\documentclass[a4paper,12pt]{article}
\usepackage{amsmath}
\usepackage{amsbsy}
\usepackage{amsthm}
\usepackage{amsfonts}
\usepackage{amssymb}
\usepackage{tikz}
\usepackage{pgfplots}
\usepackage{subfig}
\usepackage{epstopdf}
\usetikzlibrary{fit}
\usetikzlibrary{shapes.geometric}
\usetikzlibrary{decorations.pathmorphing}

\newtheorem{theorem}{Theorem}[section]
\newtheorem{lemma}{Lemma}

\numberwithin{equation}{section}

\let\a=\alpha \let\b=\beta    \let\g=\gamma     \let\d=\delta     \let\e=\varepsilon
\let\z=\zeta  \let\h=\eta     \let\th=\theta \let\k=\kappa     \let\l=\lambda
    \let\n=\nu      \let\x=\xi        \let\p=\pi        
\let\s=\sigma \let\t=\tau     \let\f=\varphi       \let\c=\chi
   \let\o=\omega     
\let\G=\Gamma        \let\L=\Lambda

\def\ss{{\underline\s}}

\def\VV{{\cal V}}\def\RR{{\cal R}}\def\LL{{\cal L}}\def\NN{{\cal N}}\def\BB{{\cal B}}\def\DD{{\cal D}}
\def\EE{{\cal E}}\def\TT{{\cal T}}\def\SS{{\cal S}}

\def\ZZZ{{\mathbb Z}}

\let\dpr=\partial
\let\io=\infty
\let\==\equiv

\def\media#1{{\left\langle#1\right\rangle}}

\def\lis{\overline}

\def\V#1{{\bf #1}}
\def\xx{{\bf x}}
\def\yy{{\bf y}}\def\kk{{\bf k}}

\def\be{\begin{equation}}
\def\ee{\end{equation}}
\def\bea{\begin{eqnarray}}\def\eea{\end{eqnarray}}

\def\pref#1{(\ref{#1})}

 \DeclareMathOperator{\supp}{supp}

\def\bml{\begin{multline}}
\def\eml{\end{multline}}

\tikzset{vertex/.style={circle,fill=black,inner sep=2pt},
bigvertex/.style={circle,fill=black,inner sep=4pt},
specialEP/.style={rectangle,fill=white,draw,inner sep=3pt},
nuEP/.style={circle,fill=white,draw, inner sep=2pt},
linelabel/.style={sloped,above,very near start, inner
sep=1pt,execute at begin node=$\scriptstyle,execute at end
node=$}, baseline=(current  bounding
box.center),doubled/.style={double distance= 1pt,line width=1.5pt}
} \pgfdeclarelayer{background} \pgfsetlayers{background,main}

\begin{document}

\title{Universal
finite size corrections and the central charge in non solvable
Ising models}

\author{Alessandro Giuliani\\
\small Dipartimento di Matematica\\ 
\small Universit\`a di Roma Tre,\\
\small L.go S. L. Murialdo 1, 00146 Roma - Italy
\and
 Vieri Mastropietro\\
\small Dipartimento di Matematica ``Federigo Enriques"\\
\small Universit\`a degli Studi di Milano\\
\small Via Cesare Saldini 50, 20133 Milano - Italy}

\date{}

\maketitle

\begin{abstract}
We investigate a non solvable two-dimensional ferromagnetic 
Ising model with nearest neighbor plus weak finite range
interactions of strength $\l$. We rigorously establish one of the 
predictions of Conformal Field Theory (CFT), namely the fact that at the critical temperature 
the finite size corrections to the free energy 
are universal, in the sense that they are exactly independent of the interaction. The 
corresponding central charge, defined in terms of the coefficient of the first subleading term to the free energy,
as proposed by Affleck and Blote-Cardy-Nightingale, is constant and equal to $1/2$
for all $0\le \l\le \l_0$ and $\l_0$ a small but finite convergence radius. 
This is one of the very few cases where the predictions of CFT can be rigorously
verified starting from a microscopic non solvable statistical model. The proof 
uses a combination of rigorous renormalization group methods with 
a novel partition function inequality, valid for ferromagnetic interactions. 
\end{abstract}

\newpage

\tableofcontents

\section{Introduction and main results}

The applications of Conformal Field Theory (CFT) to statistical mechanics 
are based on the assumption that a statistical model at the critical point 
admits a non-trivial, conformally invariant, scaling limit, 
as suggested by the renormalization theory of critical phenomena.
The two-dimensional (2D) local scale invariance 
strongly constraints the structure of the critical theory, as 
understood by Belavin, Polyakov and Zamolodchikov \cite{BPZ}. 
They recognized that the theory is characterized by 
a dimensionless constant $c$, the central charge, which is associated with
an anomaly term in the commutation relations of the stress energy tensor.
The central charge can be also defined in terms of the finite size 
corrections to the free energy at criticality 
\cite{A,BCN}.
In some cases, the critical theory is fully characterized by the 
value of $c$, which takes the form $c=1-6/m(m+1)$, $m=2,3,4,\ldots$
Once $c$ is fixed to one of these special values, the critical exponents are 
all explicitly known in terms of the Kac formula \cite{FQS}. 

In practice, the identification of 
the critical theory associated with a given microscopic lattice model is
done by inspection, 
by trying to match the known informations about the lattice model's
exponents with the Kac formula. Once a correspondence is 
established or guessed, a large number of non trivial predictions on the 
model's correlation functions at criticality can be inferred, which in general cannot be analytically derived 
by other means. It is therefore important to check these predictions in specific models, which could serve 
as benchmarks for this scheme. 
Unfortunately, there are just
a few cases, based on exactly solvable lattice models,
in which so far this correspondence could be rigorously established. 
A remarkable example is the nearest neighbor Ising model at the critical point, whose understanding 
in the scaling limit improved substantially in the last few months \cite{CHI,CS,D}.
More in general, it is very hard to rigorously compute critical exponents, correlation functions 
or finite size corrections to thermodynamic functions at the critical point. 
In recent times new methods for the analysis of non-integrable 2D spin systems, 
based on the Renormalization Group, have been developed, starting from \cite{PS} and \cite{M04},
where the authors computed the critical exponent of the so-called energy field operator
in a class of perturbed Ising
models (such as the one considered in this paper) and in a class of two stacked 
interacting Ising models  (including the 8 vertex and the Ashkin-Teller models), respectively. 
By these methods one
can try to verify some of the CFT predictions in the context of non solvable 
lattice models. 

In particular in this paper we
consider an Ising model with a generic ferromagnetic short range interaction,
with Hamiltonian
\be H=-J\sum_{\xx\in
\L_{\ell,L}}\sum_{j=1,2}\sigma_{\xx}\s_{\xx+\hat{\bf
e}_j}-\l\sum_{\{\xx,\yy\}}\s_\xx v(\xx-\yy)\s_\yy\;,\label{2.1}\ee
where $J$ and $\l$ are positive constants, $\L_{\ell,L}\subset {\mathbb
Z}^2$ is a finite rectangular box of sides $L$ and $\ell$ with
periodic boundary conditions, $\s_\xx\in\{\pm1\}$, and $\hat{\bf
e}_j$ are the two unit coordinate vectors on $\ZZZ^2$. The sum in
the second term of Eq.\pref{2.1}  is over all unordered pairs of
sites in $\L_{\ell,L}$; the interaction potential $v(\xx-\yy)$ is rotation
invariant, positive and has finite range, namely: $v(\xx-\yy)=0$, $\forall
|\xx|>R_0:= M_0$, for a suitable positive integer $M_0$. With no loss of generality, we can assume 
that $v(\V0)=v({\bf e}_j)=0$. The case
$\l=0$ corresponds to the nearest-neighbor Ising model which is
exactly solvable \cite{On44,Ka49, KO49, Ya52,MW}; in the case
$\l\not=0$ no solution is known but in \cite{PS} the exponent of
the energy-energy correlation was computed and shown to be {\it
universal} (i.e., equal to $2$ as in the $\l=0$ case), in
contrast with the critical temperature or the amplitude of the
correlations which are model dependent.

A number of key informations on the system are encoded in the
{\it partition function}:
\be \mathcal Z_\b(\L_{\ell,L})= \sum_{\ss\in \L_{\ell,L}} e^{-\b H(\ss)}\;. \label{0.1}\ee
At all temperatures, the thermodynamic limit for the {\it pressure} is well-defined
and independent of the speed at which $\ell$ and $L$ are sent to infinity:
\be p_\b=\lim_{\ell,L\to\io}\frac1{ \ell L}\log \mathcal Z_\b(\L_{\ell,L})\;.\ee
While the mere existence of the limit can be proved very generally, based on convexity and 
subadditivity arguments \cite{Ru69}, the explicit form of $p_\b$ can be computed either by
the Onsager's solution (for $\l=0$) or by cluster expansion and Renormalization Group analysis 
(for $\l$ sufficiently small, see below). It turns out that $p_\b$ depends explicitly on $\l$, at all
temperatures $\b$. Moreover, at all temperatures but the critical one, the limit is reached 
exponentially fast. At $\b=\b_c(\l)$ the limit is reached polynomially, and we shall define
$f_\io:=p_{\b_c(\l)}$.  
Remarkably, according to the ideas and methods of CFT, the finite size corrections to 
$f_\io$ are expected to be universal, in particular independent of $\l$ and $v$. More
precisely, in the presence of
periodic boundary conditions, Ref.\cite{A,BCN} predicted the validity of the following formula,
asymptotically for large $\ell$:
\be \lim_{L\to\infty}\frac{\log \mathcal Z_{\b_c(\l)}(\L_{\ell,L})} {\ell
L}=f_{\io}+\frac{c\p}{6}\frac1{\ell^2}+o(\frac1{\ell^2})\;.\label{a.11}\ee
where $c$ is the central charge of the critical theory. For the nearest neighbor Ising model $c=1/2$
and supposedly the same should be true for perturbed Ising models of the form \pref{2.1}.
Our main results is a rigorous confirmation of this expectation.
\begin{theorem}\label{thm1} Given the model \pref{2.1} with $\l$ positive and small enough,
there exists a critical temperature $\b_c(\l)$
such that Eq.\pref{a.11} holds with \be c=1/2\label{c12}\ee
\end{theorem}
Recall that the critical exponent of the energy field operator 
is known to be universal and equal to 2  \cite{PS,GGM}, as expected from the 
use of the Kac formula at $c=1/2$ (see e.g. \cite{FQS}): therefore, our result says not only that the 
finite size corrections to the free energy are universal at criticality and that the corresponding central charge is 
constant, but also that such central charge has the {\it right value}, i.e. the one matching with the one 
guessed from the critical exponents and the Kac formula. This is a very non trivial connection, predicted 
by CFT, between critical exponents and finite size corrections to the thermodynamic functions, and our results is the 
first proof of the correctness of this prediction in a non-integrable statistical model. It would be very interesting to 
extend the connection to the critical exponent of the spin field operator, but this seems a much harder problem than the 
one solved here; on the basis of the above correspondence, such exponent is expected to be $1/4$, but its 
rigorous computation at $\l\neq 0$ is still beyond reach of the current 
techniques. Another very interesting extension would be to prove an analogous universality result 
for statistical models with $c=1$, such as the 8 vertex or the Ashkin-Teller model. 
In these cases several critical exponents have been computed and proved to be $\l$-dependent 
\cite{M04,GM05, BFM09, BFM10}, as expected from CFT; at the same time, the finite size correction to the free energy 
are expected to be independent of $\l$ and given by  Eq.(\ref{a.11}) with $c=1$, but this fact is unproved so far. 
We hope to come back to this question in a future publication. 

In the case of the nearest-neighbor Ising model, $\l=0$, the result Eq.(\ref{c12}) was
proved by \cite{FF} and follows from the exact
solution. If $\l\neq 0$, the proof of our main theorem uses a Renormalization
Group (RG) analysis first introduced in \cite{PS} for the computation of the two-point 
energy correlation function, and recently extended in \cite{GGM}
to the analysis of the $n$-point energy correlations. In addition to the ideas of 
\cite{PS,GGM}, we use here some novel partition function inequalities, which we can only prove
for ferromagnetic interactions, $\l\ge0$.
The assumption of positivity of the interaction
is expected to be technical and we believe that the analogue of Theorem \ref{thm1} should hold
also for $\l$ small and negative. 

More in detail, our proof proceeds as follows.
We start by writing the partition function of the interacting Ising model
as a sum of four Grassmann integrals with different boundary
conditions, in a way that naturally extends the analogous 
representation for the nearest neighbor model.
Using a partition function inequality, we reduce ourselves to the study of just one out of these 
four Grassmann partition functions, namely the one with antiperiodic boundary conditions; 
we prove that the other three terms are
subleading and do not contribute to
the central charge. On the other side, the Grassmann partition function with antiperiodic boundary conditions 
can be rewritten as the product of two terms: the first is equal to the non-interacting Grassmann partition 
function with renormalized 
parameters (this is the contribution from the ``infrared fixed point"), 
while the rest includes all the corrections coming from finite infrared scales and from the irrelevant terms. Now, remarkably:
\begin{enumerate}
\item 
the dependence of the first factor upon the renormalized parameters can be scaled out by 
a simple change of variables, after which the factor takes the form $Z^{2\times Volume}\times$(free partition function), 
with $Z$ the ``wave function renormalization", which is volume-independent; therefore, the presence of the renormalized
parameters changes the bulk pressure, but not the finite size corrections;
\item the rest can be studied by a multiscale analysis, which requires the introduction of two 
running coupling constants, playing the role of wave function and critical temperature counterterms; 
these running coupling constant go to zero exponentially fast in the infrared limit, thanks to a dimensional improvement in the dimensional bounds following from the fact that the theory is super-renormalizable, in the sense that all the 
field operators with more than two fields are irrelevant in the RG sense; correspondingly, the finite size corrections to the pressure coming from these term can be shown to go to zero faster than $\ell^{-2}$ as 
the infrared cutoff $\ell$ is removed, namely like $\ell^{-2-\th}$, for $0<\th<1$.
\end{enumerate}
The strategy resembles closely the one used in \cite{GMP1,GMP2} to prove the universality of the optical conductivity 
in interacting graphene. An extra difficulty that we have to face in our case is the definition of the localization and renormalization procedure at finite volume, which uses and extends the strategy proposed in \cite{BM01}.

The rest of the paper is organized as follows. In Section \ref{sec22} we review the proof of Theorem \ref{thm1}
in the non-interacting case $\l=0$. In Section \ref{sec33} we prove the main theorem in the $\l\neq 0$ case: we first 
review the Grassmann representation of the interacting partition function (Section \ref{sec3.1});
then we state the partition function inequality that we use to effectively eliminate three out of the four Grassmann 
partition function
(Section \ref{sec3.2}); next we describe 
the RG computation of the antiperiodic Grassmann partition function (Section \ref{subsec3.3}), and 
we use it to compute the bulk and subleading contributions to the pressure (Section \ref{subsec3.4}). 
In Section \ref{sec4} we prove the partition function inequality stated in Section \ref{sec3.2}. In the Appendix, 
we study a subleading correction to the pressure coming from the ratio of the Grassmann partition functions, 
both in the non-interacting (Appendix \ref{seca.1}) and in the interacting case (Appendix \ref{subsec3.4}).

\section{The nearest-neighbor Ising model}\label{sec22}

In this section we review the proof of our main theorem in the case of the nearest neighbor (n.n.) Ising model,
$\l=0$. The proof 
can be found in \cite{FF}, but is reproduced here for the reader's convenience. 
From now on we shall drop the dependence on $\b$ in the symbol used for the partition 
function, in order to avoid a too cumbersome notation. We shall be as explicit as possible in 
distinguishing the formulas where the value of $\b$ is generic from those where $\b$ is fixed to be the critical one. We shall also assume that $\ell$ and $L$ are even, in order to simplify the signs
appearing in some formulas. 

The starting point is the representation of the n.n. Ising model's partition function $\mathcal Z^0(\{J_b\};\L_{\ell,L})$ with periodic boundary 
conditions and bond-dependent link variables in terms of a sum over multipolygons (see e.g. \cite{Ga}):
\be \mathcal Z^0(\{J_b\};\L_{\ell,L})=2^{\ell L}\big[\prod_{b\in \mathcal B_{\ell,L}}\cosh(\b J_b)\big] \sum_{\G\subseteq \L_{\ell,L}}\prod_{\g\in\G} \prod_{b\in \g}\tanh(\b J_b)\;,\label{h13}
\ee
where $\mathcal B_{\ell,L}$ is the set of n.n. bonds in $\L_{\ell,L}$ and $\G$ is a collection of disjoint polygons 
in $\L_{\ell,L}$,  each of which is 
a closed connected collection of bonds; here closed means that at each of the vertices covered by the polygon 
there is an even number (either 2 or 4) of incident bonds. Note that some of the polygons in $\G$ can wind up the torus 
$\L_{\ell,L}$, due to the periodic boundary conditions. Note also that the case $\G=\emptyset$ is included in the sum, in which case the corresponding contribution is equal to 1.

A convenient (for computational purposes) way of re-expressing the partition sum Eq.(\ref{h13}) is by 
writing it in terms of Grassmann integrals, see e.g.  \cite{S80a,GM05,GGM}. Define
\bea&&\mathcal Z^0_{\boldsymbol{\alpha}}(\{J_b\};\L_{\ell,L})=2^{\ell L}\big[\prod_{b\in \mathcal B_{\ell,L}}\cosh(\b J_b)\big]
\int \DD\Phi\,e^{S(\{J_b\};\Phi)}\;, \nonumber\\
 &&
S(\{J_b\};\Phi)=\sum_{\xx\in \L}\Big[\tanh(\b J_{(\xx,\xx+\hat {\bf e}_1)})\lis H_\xx H_{\xx+\hat{\bf
 e}_1}+\tanh(\b J_{(\xx,\xx+\hat {\bf e}_2)})\lis V_\xx V_{\xx+\hat{\bf
 e}_2}\nonumber\\
&&+\lis H_\xx H_\xx +\lis V_\xx V_\xx+\lis V_\xx \lis H_\xx+V_\xx \lis H_\xx+ H_\xx \lis V_\xx+ V_\xx H_\xx\Big]\;.\label{2.5tt}%
\eea
Here $\lis H_\xx, H_\xx,\lis V_\xx,  V_\xx$ are independent {\it
Grassmann variables}, four for each lattice site, $\Phi=\{\lis
H_\xx,H_\xx,\lis V_\xx,$ $V_\xx\}_{\xx\in \L}$ denotes the
collection of all of these Grassmann symbols and $ \DD \Phi$ is a
shorthand for $\prod_{\xx}d\lis H_\xx dH_\xx d\lis V_\xx dV_\xx$.
The label $\boldsymbol\a=(\a_1,\a_2)$, with $\a_1,\a_2\in\{\pm\}$,
refers to the boundary conditions, which are periodic or
antiperiodic in the horizontal (resp. vertical) direction,
depending on whether $\a_1$ (resp. $\a_2$) is equal to $+$ or $-$.
The connection between the Grassmann integral and the mutlipolygon representation can be made apparent 
by expanding the exponential inside the Grassmann integral and by integrating term by term. The result is
\bea && \mathcal Z^0_{\boldsymbol{\alpha}}(\{J_b\};\L_{\ell,L})=2^{\ell L}\big[\prod_{b\in \mathcal B_{\ell,L}}\cosh(\b J_b)\big]\cdot\label{gr}\\
&&\qquad\cdot\sum_{\G\subseteq \L_{\ell,L}}(-\a_1)^{h(\G)}(-\a_2)^{v(\G)}(-1)^{h(\G)v(\G)}\prod_{\g\in\G} 
\prod_{b\in\g}\tanh(\b J_b)\;,
\nonumber\eea
where $h(\G)$ and $v(\G)$ are the number of windings of $\G$ on the torus, in the horizontal and vertical directions,
respectively. The r.h.s. of Eq.(\ref{gr}) is very similar to the multipolygon representation Eq.(\ref{h13}), modulo 
the sign $(-\a_1)^{h(\G)}(-\a_2)^{v(\G)}$ $(-1)^{h(\G)v(\G)}$, which depends on 
whether the parity of the number of windings in the horizontal-vertical directions are even-even, or even-odd,
or odd-even, or odd-odd. The value taken by this sign with different boundary conditions and different 
winding parities can be conveniently summarized in the following table.
%
\begin{center}
\begin{tabular}{|l|c|c|c|c|}
\hline
& even-even&even-odd&odd-even&odd-odd\\
\hline
$\boldsymbol{\alpha}=(+,+)$&+&$-$&$-$&$-$\\
\hline
$\boldsymbol{\alpha}=(+,-)$&+&+&$-$&+\\
\hline
$\boldsymbol{\alpha}=(-,+)$&+&$-$&+&+\\
\hline
$\boldsymbol{\alpha}=(-,-)$&+&+&+&$-$\\
\hline
\end{tabular}
\end{center}
More explicitly, we can write:
\bea && \mathcal{Z}^0_{++}(\{J_b\};\L_{\ell,L})=\mathcal{Z}^0_{e-e}(\{J_b\};\L_{\ell,L})-\label{4.1k}\\
&&
-\mathcal{Z}^0_{e-o}(\{J_b\};\L_{\ell,L})-\mathcal{Z}^0_{o-e}(\{J_b\};\L_{\ell,L})-\mathcal{Z}^0_{o-o}(\{J_b\};\L_{\ell,L})\;,
\nonumber\eea
\bea && \mathcal{Z}^0_{+-}(\{J_b\};\L_{\ell,L})=
\mathcal{Z}^0_{e-e}(\{J_b\};\L_{\ell,L})+\\
&&+\mathcal{Z}^0_{e-o}(\{J_b\};\L_{\ell,L})-\mathcal{Z}^0_{o-e}(\{J_b\};\L_{\ell,L})+\mathcal{Z}^0_{o-o}(\{J_b\};\L_{\ell,L})\;,
\nonumber\eea
\bea && \mathcal{Z}^0_{-+}(\{J_b\};\L_{\ell,L})=
\mathcal{Z}^0_{e-e}(\{J_b\};\L_{\ell,L})-\\
&&-\mathcal{Z}^0_{e-o}(\{J_b\};\L_{\ell,L})+\mathcal{Z}^0_{o-e}(\{J_b\};\L_{\ell,L})+\mathcal{Z}^0_{o-o}(\{J_b\};\L_{\ell,L})\;,
\nonumber\eea
\bea && \mathcal{Z}^0_{--}(\{J_b\};\L_{\ell,L})=
\mathcal{Z}^0_{e-e}(\{J_b\};\L_{\ell,L})+\label{4.4k}\\
&&+\mathcal{Z}^0_{e-o}(\{J_b\};\L_{\ell,L})+\mathcal{Z}^0_{o-e}(\{J_b\};\L_{\ell,L})-\mathcal{Z}^0_{o-o}(\{J_b\};\L_{\ell,L})\;,
\nonumber\eea
where 
\be \mathcal Z^0_{e-e}(\{J_b\};\L_{\ell,L})=2^{\ell L}\big[\prod_{b\in \mathcal B_{\ell,L}}\cosh(\b J_b)\big] \sum_{\G\subseteq \L_{\ell,L}}\nolimits^{(e-e)}\prod_{\g\in\G} \prod_{b\in \g}\tanh(\b J_b)\;,\label{h13z}
\ee
and the superscript $(e-e)$ on the sum $\sum_{\G\subseteq\L_{\ell,L}}^{(e-e)}$ 
indicates the constraint that $\G$ winds over $\L_{\ell,L}$ an even/even number of times in the horizontal/vertical direction, {\it including the case that $\G$ does not wind over the torus}; in other words, 
when we say ``even number of windings", we include the case of zero windings. Of course, the other partition functions,  namely $\mathcal Z^0_{e-o}(\{J_b\};\L_{\ell,L})$, 
$\mathcal Z^0_{o-e}(\{J_b\};\L_{\ell,L})$,
$\mathcal Z^0_{o-o}(\{J_b\};\L_{\ell,L})$, are defined similarly, with the constraint that $\G$ winds up 
over the torus an even/even number of times in the horizontal/vertical direction replaced by the one that $\G$ winds up 
an even/odd, odd/even, odd/odd number of times, respectively. By definition, 
the total partition function in Eq.(\ref{h13}) is
\bea &&  \mathcal{Z}^0(\{J_b\};\L_{\ell,L})=
\mathcal{Z}^0_{e-e}(\{J_b\};\L_{\ell,L})+\nonumber\\
&&+\mathcal{Z}^0_{e-o}(\{J_b\};\L_{\ell,L})+\mathcal{Z}^0_{o-e}(\{J_b\};\L_{\ell,L})+\mathcal{Z}^0_{o-o}(\{J_b\};\L_{\ell,L})\;.\label{4.compk}\eea
Alternatively, using Eqs.(\ref{4.1k})--(\ref{4.4k}), we can also write 
\bea &&
\mathcal Z^0(\{J_b\};\L_{\ell,L})=\frac12\big(\mathcal Z^0_{--}(\{J_b\};\L_{\ell,L})+\label{aaaa}\\
&&\qquad+\mathcal Z^0_{-+}(\{J_b\};\L_{\ell,L})+
\mathcal Z^0_{+-}(\{J_b\};\L_{\ell,L})- \mathcal Z^0_{++}(\{J_b\};\L_{\ell,L})\big)\;,\nonumber
\eea
which is the desired connection between the multipolygon and the Grassmann representations. 
This relation is valid for all $\{J_b\}_{b\in\mathcal B_{\ell,L}}$ and all inverse temperatures $\b$. 
On the other hand, if $J_b$ is independent of $b$, 
the identity Eq.(\ref{aaaa}) gives us a mean to compute the partition function in closed 
form, simply because the Grassmann integrals $\mathcal Z^0_{\boldsymbol{\alpha}}(\L_{\ell,L}):=
\mathcal Z^0_{\boldsymbol{\alpha}}(\{J_b\};\L_{\ell,L})\big|_{J_b\equiv J}$ are 
gaussian and translation invariant. In practice, the computation proceeds as follows: 
one first goes to Fourier space, thus block-diagonalizing the quadratic action appearing in the 
Grassmann integral; each block one is left with involves the degrees of freedom associated with the Fourier
modes $\kk$ and $-\kk$, with $\kk\in\mathcal D_{\boldsymbol\alpha}$, and 
\be\DD_{\boldsymbol{\alpha}}=
\Big\{\kk=\Big(\frac{2\p}{\ell}(r+\small{\frac{1-\a_1}{4}}),\frac{2\p}{L}(n+\small{\frac{1-\a_2}{4}})\Big):
\ r=0,\ldots, \ell-1;\ n=0,\ldots,L-1\Big\}\;.\ee
The computation of the Grassmann integral 
of the variables associated with each block is elementary and leads to a determinant or to a Pfaffian, depending on whether 
$\kk$ differs from $-\kk$ or not. The result is the following. If $\boldsymbol{\alpha}\neq(+,+)$,
then, defining $t=\tanh(\b J)$ and $S_t(\Phi):=S(\{J_b\};\Phi)\big|_{J_b\equiv J}$, 
\be \frac{\mathcal Z^0_{\boldsymbol{\alpha}}(\L_{\ell,L})}{\big(2\cosh^2(\b J)\big)^{\ell L}}=\int \mathcal D\Phi e^{S_t(\Phi)}=
\prod_{\kk\in \mathcal D_{\boldsymbol\alpha}} [(1+t^2)^2-2
t(1-t^2)(\cos k_1+\cos k_2)]^{1/2}\;. \ee
If, on the contrary, $\boldsymbol{\alpha}=(+,+)$,
\bea && \frac{\mathcal Z^0_{++}(\L_{\ell,L})}{\big(2\cosh^2(\b J)\big)^{\ell L}}=\int \mathcal D\Phi e^{S_t(\Phi)}=(2-(1+t)^2)(2-(1-t)^2)\cdot\nonumber\\
&&\qquad \cdot
\prod_{\substack{\kk\in \mathcal D_{++}:\\ \kk\neq{\bf 0},(\p,\p)}} [(1+t^2)^2-2
t(1-t^2)(\cos k_1+\cos k_2)]^{1/2}\;,\label{++} \eea
which is positive for $\b<\b_c$, negative for $\b>\b_c$ and vanishes at the critical point 
(the critical point $\b_c$ is defined by the condition that $t=\sqrt2-1$). The difference in the results
obtained for $\boldsymbol\alpha\neq(+,+)$ or $\boldsymbol\alpha=(+,+)$ is due to the fact that 
in the 
first case all the modes $\kk\in\mathcal D_{\boldsymbol\alpha}$ can be grouped into pairs $(\kk,-\kk)$ and, 
correspondingly, the evaluation of the gaussian Grassmann integral reduces purely to a product over determinants
(each determinant being the integral over the variables of the modes $\kk$ and $-\kk$). In the second case, all modes but two can be grouped into pairs, the two special momenta being $\kk=\V0$ and 
$\kk=(\p,\p)$; therefore the result is equal to the product of the determinants associated with the paired momenta
times the two Pfaffians coming from the modes $\V0$ and $(\p,\p)$, which give the factor $(2-(1+t)^2)(2-(1-t)^2)$
in the r.h.s. of Eq.(\ref{++}).

In conclusion, evaluating Eq.(\ref{aaaa}) at $\b_c$ in the translation invariant case 
and using the fact that the $(+,+)$ Grassmann partition function vanishes at criticality, we find:
 \bea \mathcal Z^0(\L_{\ell,L})\Big|_{\b=\b_c}&=&\frac12\big(\mathcal Z^0_{--}(\L_{\ell,L})+\mathcal Z^0_{-+}(\L_{\ell,L})+
\mathcal Z^0_{+-}(\L_{\ell,L}))\Big|_{\b=\b_c}\label{zzaa}\\
&=&\mathcal Z^0_{--}(\L_{\ell,L})\Big[\frac12\big(1+\frac{\mathcal Z^0_{-+}(\L_{\ell,L})}{\mathcal Z^0_{--}(\L_{\ell,L})}+\frac{\mathcal Z^0_{+-}(\L_{\ell,L})}{\mathcal Z^0_{--}(\L_{\ell,L})}\big)\Big]\Big|_{\b=\b_c}\;,\nonumber\eea
with
\bea && \mathcal Z^0_{--}(\L_{\ell,L})\Big|_{\b=\b_c}=(\sqrt2)^{\ell L}\prod_{\kk\in\DD_{--}}(4-2\cos k_1-2\cos k_2)^{1/2}
\nonumber\\
&& 
\mathcal Z^0_{-+}(\L_{\ell,L})\Big|_{\b=\b_c}=(\sqrt2)^{\ell L}\prod_{\kk\in\DD_{-+}}(4-2\cos k_1-2\cos k_2)^{1/2}
\label{pc}\\
&&
\mathcal Z^0_{+-}(\L_{\ell,L})\Big|_{\b=\b_c}=(\sqrt2)^{\ell L}\prod_{\kk\in\DD_{+-}}(4-2\cos k_1-2\cos k_2)^{1/2}\;.
\nonumber\eea
Taking the logarithm at both sides of Eq.(\ref{zzaa}), dividing by the volume, and taking the infinite volume limit, 
we get the bulk term $f_\io$, which is given by Onsager's formula:
\bea f_\io&=&\lim_{\ell,L\to \infty}\frac1{\ell L}\log \mathcal Z^0(\L_{\ell,L})\Big|_{\b=\b_c}\\
&=&\frac12\log 2+\frac12\int\limits_{[-\p,\p]^2}\!\!\!\frac{d\kk}{(2\p)^2}\log(4-2\cos k_1-2\cos k_2)\;.\nonumber\eea
Using the notation of Eq.(\ref{a.11}), we write the first finite volume correction to the critical pressure 
in the form $\frac{c\p}{6\ell^2}$, with 
\bea \frac{c\p}{6}&=&\lim_{\ell\to\infty}\lim_{L\to\io}\Big[\frac{\ell}{2L}
\sum_{\kk\in\mathcal D_{--}}\log(4-2\cos k_1-2\cos k_2)-\ell^2\big(f_\io-\frac12\log 2\big)\Big]\nonumber\\
&+&\lim_{\ell\to\infty}\lim_{L\to\io}\frac{\ell}{L}\log \Big[\frac12
\big(1+\frac{\mathcal Z^0_{-+}(\L_{\ell,L})}{\mathcal Z^0_{--}(\L_{\ell,L})}+\frac{\mathcal Z^0_{+-}(\L_{\ell,L})}{\mathcal Z^0_{--}(\L_{\ell,L})}\big)\Big|_{\b=\b_c}
\Big]\;.\label{sl}\eea
provided this limit exists and is finite. In Appendix \ref{seca.1} we show that the limit in the second line is equal to zero. On the contrary,
the one in the first line is non trivial and can be explicitly computed as follows. Taking the limit $L\to\io$ first, we 
can rewrite the first line as
\be \lim_{\ell\to\infty}\ell^2\sum_{n=0}^{\ell/2-1}
\int_{\x_n-\frac{\pi}{\ell}}^{\x_n+\frac{\pi}{\ell}}\frac{dk_1}{2\p}\int_{-\pi}^\pi\frac{dk_2}{2\p}
\big[\log(4-2\cos \x_n-2\cos k_2)-\log(4-2\cos k_1-2\cos
k_2)\big]\;,\ee
where $\x_n=\frac{2\p}{\ell}(n+\frac12)$. The integral over $k_2$ can be performed explicitly \cite[Formula 4.224(9)]{GR}, leading to
\be  \lim_{\ell\to\infty}\ell^2\sum_{n=0}^{\ell/2-1}
\int_{\x_n-\frac{\pi}{\ell}}^{\x_n+\frac{\pi}{\ell}}\frac{dk_1}{2\p}\big[\g(\x_n)-\g(k_1)\big]=
\lim_{\ell\to\infty}\ell^2\sum_{n=0}^{\ell/2-1}
\int_{-\frac{\pi}{\ell}}^{\frac{\pi}{\ell}}\frac{dk'}{2\p}\big[\g(\x_n)-\g(\x_n+k')\big]
\;,\label{zsd}\ee
with $\g(k):=\cosh^{-1}(2-\cos k)$. Expanding in Taylor series $\g(\x_n+k')$ around $\x_n$ up to second order included, 
and using the fact that $\g(k)$ is a $C^\io$ function on $[0,\pi]$, we find that Eq.(\ref{zsd}) can be rewritten as
\bea && \lim_{\ell\to\infty}\Big[-\frac{\ell^2}2\sum_{n=0}^{\ell/2-1}\g''(\x_n)\int_{-\frac{\pi}{\ell}}^{\frac{\pi}{\ell}}\frac{dk'}{2\p}(k')^2
+O(\frac1\ell)\Big]=\\
&&=-\frac{\p}{12}\int_{0}^\p dk\, \g''(k)=\frac{\p}{12}(\g'(0^+)-\g'(\p))=\frac{\p}{12}\equiv \frac{c\p}{6}
\;,\nonumber\eea
which corresponds to $c=1/2$, as desired. 

\section{The interacting case}\label{sec33}

We now attack the problem of computing the first non trivial finite volume correction to the critical 
pressure in the interacting, $\l\neq0$, case. We make use of the results and methods of \cite{GGM}, which  
we refer to for the proof of numerous relations used in the following. As in the previous section,
we drop the dependence on $\b$ in the symbol used for the partition 
function, and we assume that $\ell$ and $L$ are even.

\subsection{Grassmann representation of the interacting partition function}\label{sec3.1}

The interacting partition function Eq.(\ref{0.1}) 
can be written in a form analogous to Eq.(\ref{aaaa}), for all temperatures $\b$:
\be \mathcal Z(\L_{\ell,L})=\frac12\big(\mathcal Z_{--}(\L_{\ell,L})+\mathcal Z_{-+}(\L_{\ell,L})+ 
\mathcal Z_{+-}(\L_{\ell,L})- \mathcal Z_{++}(\L_{\ell,L})\big)\;,\label{aaaa11} \ee
with $\mathcal Z_{\boldsymbol\alpha}(\L_{\ell,L})$ given by (see  
\cite[Proposition 1]{GGM}):
\be \mathcal Z_{\boldsymbol\alpha}(\L_{\ell,L})=C_{\ell,L}\int \mathcal D\Phi\, e^{S_t(\Phi)+\mathcal V(\Phi)}\;,
\label{prop1}
\ee
where:
\begin{itemize}
\item $C_{\ell,L}$ is a normalization constant, defined as
\be 
C_{\ell,L}=(2\cosh^2(\b J))^{\ell L}e^{V_{\ell,L}(\l)}\prod_{\{\xx,\yy\}} \cosh^2\!\big(\frac{\b\l}2 v(\xx-\yy)\big) \label{2.52}
\ee
with $V_{\ell,L}(\l)$ an analytic function of $\l$, defined as (using the notation of \cite{GGM}, see 
the proof of \cite[Proposition 1]{GGM} and, in particular, \cite[Eq.(2.29)]{GGM})
\be V_{\ell,L}(\l)=2\ell L\sum_{\substack{\G\subseteq\L_{\ell,L}:\\
\supp\G\ni b_0}} \frac{\f^T(\G)}
{|\supp\G|} \prod_{\g\in\G} \z(\emptyset,\emptyset;\g)\;.\label{forV}\ee
Here $b_0$ is an arbitrary n.n. bond of $\L_{\ell,L}$, $\supp\G=\cup_{\g\in\G}\g$ and $|\supp\G|$ is the number of bonds in $\supp\G$.
The sum in Eq.(\ref{forV}) is independent of $b_0$, by translation invariance; the {\it activity}
$\z(\emptyset,\emptyset;\g)$ (defined in \cite[Eq.(2.18)]{GGM}) is a translation invariant exponentially decaying 
function, satisfying the bound (see \cite[Eq.(2.28)]{GGM})
\be |\z(\emptyset,\emptyset;\g)|\le \n^{|\g|}\;,\qquad \n=4 e^{1+\b|\l|/2}\big(\frac{\b|\l|}{2}\big)^{1/M_0}\;.
\label{eq:zDef}
\ee
\item If we define $E_{\xx,1}=\lis H_\xx H_{\xx+a\hat{\bf e}_1}$ and
$E_{\xx,2}=\lis V_\xx V_{\xx+a\hat{\bf e}_2}$, then $\VV(\Phi)$ is a polynomial in $\{E_{\xx,j}\}_{\xx\in \L_{\ell,L}}^{j=1,2}$,
which can be expressed as
\be\VV(\Phi) = \sum_{n\ge 1}\sum_{j_1,\ldots,j_n}
\sum_{\xx_1,\ldots,\xx_n}W_{\underline{j}}(\xx_1,\ldots,\xx_n)
\prod_{i=1}^n E_{\xx_i,j_i}\label{V_as_W}\ee
where $\underline{j}=(j_1,\ldots,j_n)$ and
 the kernel $W_{\underline{j}}(\xx_1,\ldots,\xx_n)$ is translation invariant and satisfies the
following decay bound:
\be
|W_{\underline{j}}(\xx_1,\ldots,\xx_n)|\le 
C^{n}(\b |\l|)^{\max\{1,cn\}} e^{-\k\, \d(\xx_1,\ldots,\xx_n)}\label{W_decay}\ee
for suitable constants $C,c,\k>0$ depending only on $M_0$ (the range of the interaction); 
here $\d(\xx_1,\ldots,\xx_n)$ is the tree 
distance of the set $X=\{\xx_1,\ldots,\xx_n\}$, that is the length of the shortest tree graph composed of
bonds in $\BB_{\ell,L}$ which connects all the elements of $X$.
\end{itemize}
The Grassmann integral Eq.(\ref{prop1}) and the correlation functions of the $\Phi$ field induced by the 
``measure" $\mathcal D\Phi\, e^{S_t(\Phi)+\mathcal V(\Phi)}$ have been studied in great detail in \cite{GGM}. 
In particular, part of the main result of \cite{GGM} can be reinterpreted by saying that there 
is a critical temperature $\b_c(\l)$ such that, if we fix $t=t_c=\tanh(\b_c(\l)J)$, then the two-point function of the 
$\Phi$ field decays polynomially (like ${\rm distance}^{-1}$) at large distances: if $\boldsymbol\alpha\neq(+,+)$ and we perform the unitary  
change of variables from $\Phi$ to the {\it critical modes} $\psi,\chi$ defined as
\be
\begin{pmatrix}
{\psi}_{\xx,+} \\ \psi_{\xx,-} \\ {\chi}_{\xx,+} \\
\chi_{\xx,-}
\end{pmatrix}
= U \begin{pmatrix}
\lis H_{\xx} \\ H_{\xx} \\ \lis V_\xx \\
V_\xx
\end{pmatrix}
\;,\qquad U = \frac12
\begin{pmatrix}
e^{i\frac{\pi}4} & e^{-i\frac\pi4} & 1 & -i \\
e^{-i \frac\pi4} & e^{i\frac{\pi}4} & 1 & i \\
-e^{i\frac{\pi}4} & -e^{-i\frac\pi4} & 1 & -i \\
-e^{-i\frac\pi4} & -e^{i\frac{\pi}4} & 1 & i
\end{pmatrix}\;,\label{u1}
\ee
then, asymptotically for large distances,
\be\media{\psi_{\xx,\o}\psi_{\yy,\o'}}_{t_c}=\frac{\int \mathcal D\Phi e^{S_{t_c}(\Phi)+\mathcal V(\Phi)}\psi_{\xx,\o}\psi_{\yy,\o'}}
{\int \mathcal D\Phi e^{S_{t_c}(\Phi)+\mathcal V(\Phi)}}\simeq
\frac{\bar Z(\l)}{\p t_c}\frac{\d_{\o,\o'}}{(y_1-x_1)+i\o(y_2-x_2)}\;.\label{3.9}\ee
Here the symbol ``$\simeq$" means ``up to faster decaying terms as $|\xx-\yy|\to \io$" and 
$\bar Z(\l)$ is the analytic function 
appearing in \cite[Theorem 1.1]{GGM}. An infinite volume limit $\ell,L\to\io$, performed while keeping 
the sites $\xx,\yy$ fixed, is implicit in Eq.(\ref{3.9}). Eq.(\ref{3.9}) can be read by saying that the asymptotic 
behavior of $\media{\psi_{\xx,\o}\psi_{\yy,\o'}}_{t_c}$ is the same as that of 
$\media{\psi_{\xx,\o}\psi_{\yy,\o'}}^0_{t_c^0,Z}$, where $t_c^0=\sqrt2-1$, $Z=\frac{1}{\bar Z(\l)}\frac{t_c}{t_c^0}$ and 
$\media{\cdot}^0_{Z,t_c^0}$ is the average with respect to a properly renormalized massless gaussian integration: 
\be \media{\ \cdot\ }^0_{Z,t_c^0}=\frac{\int \mathcal D\Phi e^{Z S_{t_c^0}(\Phi)}\ \cdot\ }
{\int \mathcal D\Phi e^{Z S_{t_c^0}(\Phi)}}\;.\label{free}\ee
A way of computing $Z$, similar to but slightly different from the one proposed in \cite{GGM}, will be described
below. The integration $\mathcal D\Phi e^{Z S_{t_c^0}(\Phi)}$ is the right reference measure, around which to
perform the perturbation analysis of Eq.(\ref{prop1}) at $t=t_c$, along the lines of \cite{GGM}. \\

However, before we start describing the renormalization group computation of 
$\mathcal Z_{\boldsymbol\alpha}(\L_{\ell,L})$ at $t=t_c$, let us discuss how to
deal with the sign problem in Eq.(\ref{aaaa11}): it is in fact apparent that the expression in the r.h.s. involves 
a difference between Grassmann partition functions, which may in principle produce dangerous cancellations
between the different terms. This has to be contrasted with the computation in the $\l=0$ where, as we saw above, 
at criticality $\mathcal Z^0_{++}(\L_{\ell,L})=0$ and the other three terms are all positive. This may even be true 
at $\l\neq 0$ and $t=t_c$, but we do not know how to prove it. Nevertheless, we can prove an a priori partition 
function inequality, which can be thought of as a weak version of this claim, 
and is actually enough  to the purpose of computing the 
pressure up to the first non trivial finite volume correction. This is discussed in the next subsection.

\subsection{A partition function inequality}\label{sec3.2}

As we said above, our goal is to compute Eq.(\ref{aaaa11}) at criticality, by using the representation Eq.(\ref{prop1})
and a renormalization group analysis for $\int \mathcal D\Phi\, e^{S_t(\Phi)+\mathcal V(\Phi)}$, along the lines
of \cite{GGM}. However, as we will see below, our renormalization group computation of 
$\int \mathcal D\Phi\, e^{S_t(\Phi)+\mathcal V(\Phi)}$ only works for $\boldsymbol\alpha\neq(+,+)$, due to 
the possible vanishing of $\mathcal Z_{++}(\L_{\ell,L})$ at $t=t_c$; the problem is that at the unperturbed level 
$\mathcal Z^0_{++}(\L_{\ell,L})=0$ at criticality and, therefore, in order to perturbatively 
compute $\mathcal Z_{++}(\L_{\ell,L})$, we do not know where to perturb around. 
Luckily enough, in order to compute the pressure up to the first non trivial finite volume correction, we 
do not really need to prove that $\mathcal Z_{++}(\L_{\ell,L})$ is zero or much smaller than 
the other three partition functions: a weaker statement, summarized in the following 
Lemma, is actually enough for our purposes. 

\begin{lemma}\label{lemma1}
Under the stated assumptions on the potential $v(\xx)$ in Eq.(\ref{2.1}), for $\l\ge 0$ and 
all inverse temperatures $\b>0$, the following inequalities hold: 
\bea && \frac13\le \frac{\mathcal Z(\L_{\ell,L})}{\mathcal Z_{--}(\L_{\ell,L})+\mathcal Z_{-+}(\L_{\ell,L})+\mathcal Z_{+-}(\L_{\ell,L})}\le 1\;,\label{3.11a}\\
&&\mathcal Z_{-+}(\L_{\ell,L})+\mathcal Z_{+-}(\L_{\ell,L})\ge 0\;.\label{3.12a}\eea
\end{lemma}
The proof of this lemma is postponed to Section \ref{sec4}. In order to use this result, we have to combine it 
with the information that 
\be \mathcal Z_{--}(\L_{\ell,L})>0\;,\label{3.13a}\ee
which is proved in the next subsection for $\l$ small enough. Putting Eqs.(\ref{3.11a})--(\ref{3.13a}) together we get:
\be \frac13\mathcal Z_{--}(\L_{\ell,L})\le \mathcal Z(\L_{\ell,L})\le \mathcal Z_{--}(\L_{\ell,L})
\Big(1+\frac{|\mathcal Z_{-+}(\L_{\ell,L})|}{\mathcal Z_{--}(\L_{\ell,L})}+\frac{|\mathcal Z_{+-}(\L_{\ell,L})|}{\mathcal Z_{--}(\L_{\ell,L})}\Big)\;.\ee
Let us now put ourselves at criticality, $t=t_c$. In Appendix \ref{subsec3.5} we prove that 
\be \lim_{\ell\to\infty}\lim_{L\to\io}\frac{\ell}{L}\log \Big(1+\frac{|\mathcal Z_{-+}(\L_{\ell,L})|}{\mathcal Z_{--}(\L_{\ell,L})}+\frac{|\mathcal Z_{+-}(\L_{\ell,L})|}{\mathcal Z_{--}(\L_{\ell,L})}\Big)=0\label{3.10b}\ee
and, therefore, using the notation of Eq.(\ref{a.11})
\be f_\io=\lim_{\ell,L\to\io}\frac1{\ell L}\log \mathcal Z(\L_{\ell,L})=\lim_{\ell,L\to\io}\frac1{\ell L}\log \mathcal Z_{--}(\L_{\ell,L})\;,\label{fio}\ee
\be \frac{c\p}{6}=\lim_{\ell\to\io}\lim_{L\to\io}\Big[\frac{\ell}{L}\log \mathcal Z(\L_{\ell,L})-\ell^2f_\io\Big]
=\lim_{\ell\to\io}\lim_{L\to\io}\Big[\frac{\ell}{L}\log \mathcal Z_{--}(\L_{\ell,L})-\ell^2f_\io\Big]\;,\label{fiode}\ee
which reduces the computation of the central charge to the evaluation of the finite volume corrections to 
$\mathcal Z_{--}(\L_{\ell,L})$. These are computed in the next two subsections. 

\subsection{Renormalization group analysis of $\mathcal Z_{--}(\L_{\ell,L})$}\label{subsec3.3}

We start from Eq.(\ref{prop1}) with $\boldsymbol\alpha=(-,-)$ and $t=t_c$, and we rewrite it as:
\be \mathcal Z_{--}(\L_{\ell,L})=C_{\ell,L}\int \mathcal D\Phi\, e^{ZS_{t_c^0}(\Phi)+\lis{\mathcal V}(\Phi)}\;,
\label{3.11}
\ee
where 
\be \lis\VV(\Phi)=\VV(\Phi)+S_{t_c}(\Phi)-Z S_{t_c^0}(\Phi)\;. \label{3.18}\ee
In the following, $t_c$ and $Z$ will be constructed in such a way that, asymptotically for $|\xx|\to\io$,
\be \media{\psi_{\V0,\o}\psi_{\xx,\o'}}_{t_c}\simeq \media{\psi_{\V0,\o}\psi_{\xx,\o'}}_{t_c^0,Z}^0\simeq
\frac1{Z}\frac1{\p t_c^0}\frac{\d_{\o,\o'}}{x_1+i\o x_2}\ee
that is the theory is critical (i.e. massless) and $Z\p t_c$ is the dressed wave function renormalization, 
see the discussion after Eq.(\ref{3.9}). Equivalently, the $t_c$ and $Z$ will be chosen in such a way that 
the flow of the running coupling constants defined in the iterative construction of $\mathcal Z_{--}(\L_{\ell,L})$
described below remain bounded at all scales; we refer to the following for the definition of running coupling 
constants and for the study of their flow; see in particular the neighborood of 
Eqs.(\ref{3.70})) and (\ref{3.70bis}
for a discussion about how to fix $t_c,Z$ in terms of an implicit function theorem.

It is convenient to multiply and divide Eq.(\ref{3.11}) by the proper normalization, $\tilde{\mathcal Z}^0_{--}(\L_{\ell,L})=\int \mathcal D\Phi e^{ZS_{t_c^0}(\Phi)}$, thus finding
\be \mathcal Z_{--}(\L_{\ell,L})=C_{\ell,L}\tilde{\mathcal Z}^0_{--}(\L_{\ell,L})
\int P(d\Phi)\, e^{\lis{\mathcal V}(\Phi)}\;,
\label{3.14}\ee
where $P(d\Phi)= \big[\tilde{\mathcal Z}^0_{--}(\L_{\ell,L})\big]^{-1}\mathcal D\Phi\, e^{ZS_{t_c^0}(\Phi)}$
is the normalized gaussian reference measure. The normalization constants $C_{\ell,L}$ and 
$\tilde{\mathcal Z}^0_{--}(\L_{\ell,L})$ are explicit and, therefore, the only non trivial part to deal with is
\be \Xi_{--}(\L_{\ell,L}):=\int P(d\Phi)\, e^{\lis{\mathcal V}(\Phi)}\;.\label{3.21}\ee
In order to compute this integral we proceed by following essentially the same strategy of \cite{GGM}, modulo a few
small modifications described below. As a first step, we pass to the critical modes, already introduced 
in Eq.(\ref{u1}); in order to fix the normalizations as in \cite{GGM} we also 
rescale the variables as (see \cite[Eq.(2.38)]{GGM})
$\psi_\o\to -i\o\sqrt{\p t_c^0}\,\psi_\o$, $\c_\o\to -i\o\sqrt{\p t_c^0}\,\c_\o$, 
and next we perform the following linear change of variables (see \cite[Eq.(2.52)]{GGM}))
\be \hat \c_\kk\to
\hat\c_\kk +C_\c^{-1}(\kk)Q(\kk)\hat\psi_\kk\;,\label{2.lin}\ee
where, if $\s_\c(\kk)=\cos k_1+\cos k_2+2\frac{\sqrt 2+1}{t_c^0}$, 
\bea &&C_\c(\kk) = \begin{pmatrix}
-i \sin k_1 +\sin k_2 & i \s_\c(\kk) \\
-i\s_\c(\kk)& -i \sin k_1-\sin k_2
\end{pmatrix}\;,\label{2.90a}\\
&&Q(\kk)=
\begin{pmatrix}
-i\sin k_1- \sin k_2 & i \cos k_1 - i\cos k_2 \\
- i \cos k_1 +i\cos k_2 & -i\sin k_1+\sin k_2 \end{pmatrix}\;.\eea
After these changes of variables we get the analogue of \cite[Eq.(2.53)]{GGM}, namely
\be \Xi_{--}(\L_{\ell,L})=\int
P(d\psi)P(d\c) e^{\lis\VV(\psi,\c)}\;, \label{2.90}\ee
where:
\begin{itemize}
\item If, for $\kk\in\mathcal D_{--}$, we define 
$$\hat \psi_{\kk,\o}=\sum_{\xx\in\L_{\ell,L}}e^{i\kk\xx}\psi_{\xx,\o}\;,\qquad 
\hat \chi_{\kk,\o}=\sum_{\xx\in\L_{\ell,L}}e^{i\kk\xx}\chi_{\xx,\o}\;,$$
then the Grassmann gaussian integrations 
$
P(d\psi)$, $P(d\c)$ can be written as
\bea && 
P(d\psi):=\frac1{
\NN_{\psi,{\boldsymbol\a}}}\Big[\prod_{\kk\in\DD_{--}}\prod_{\o=\pm}d\hat\psi_{\kk,\o}\Big] \exp\Big\{- \frac{Z}{4\p L\ell} \sum_{\kk \in 
\DD_{--}} \hat\psi^T_{-\kk} 
C_\psi(\kk)\hat\psi_{\kk}\Big\}\;,\nonumber\\
&&P(d\c):=\frac1{\NN_{\c,{\boldsymbol\a}}}\Big[\prod_{\kk\in\DD_{--}}\prod_{\o=\pm}d\hat\c_{\kk,\o}\Big] \exp\Big\{- \frac{Z}{4\p L\ell} \sum_{\kk \in 
\DD_{--}} \hat\c^T_{-\kk} C_\c(\kk)\hat\c_{\kk}\Big\}\;,\nonumber\eea
where, letting $\s_\psi(\kk):=\cos k_1+\cos k_2-2$, 
\be 
C_\psi(\kk)= \begin{pmatrix}
-i \sin  k_1 +\sin  k_2 & i \s_\psi(\kk)\\
-i\s_\psi(\kk)& -i \sin  k_1-\sin  k_2\end{pmatrix}
-Q(\kk)C_\c^{-1}(\kk)Q(\kk)\ee
and $
\NN_{\psi}, \NN_{\c}$ two normalizations, such that $\int 
P(d\psi)=\int P(d\c)=1$.
\item $\lis \VV(\psi,\c)$ is the rewriting of $\lis \VV(\Phi)$ in terms of the new variables. It is easy to check that 
its kernels satisfy the same decay estimates as those of $\lis \VV(\Phi)$, see Eq.(\ref{W_decay}).
\end{itemize}
The propagators of the $\psi$ and $\c$ fields are, respectively,
\bea g^\psi_{\o,\o'}(\xx-\yy)&=&\int 
P(d\psi) \psi_{\xx,\o}\psi_{\yy,\o'}=\frac{1}{Z}
\frac{2\p}{L\ell}\sum_{\kk\in\DD_{--}}e^{-i\kk(\xx-\yy)}\big(\big[
C_\psi(\kk) \big]^{-1}\big)_{\o,\o'}\;,\nonumber\\
g^\c_{\o,\o'}(\xx-\yy)&=&\int P(d\c) \c_{\xx,\o}\c_{\yy,\o'}=\frac{1}{Z}
\frac{2\p}{L\ell}\sum_{\kk\in\DD_{--}}e^{-i\kk(\xx-\yy)}\big(\big[C_\chi(\kk) \big]^{-1}\big)_{\o,\o'}
\;.\nonumber\eea
A simple explicit computation shows that asymptotically as $|\xx-\yy|\to\io$ the propagator of the $\psi$ 
field behaves like $\frac1{Z}\frac{\d_{\o,\o'}}{(x_1-y_1)+i\o(x_2-y_2)}$, as it should, while the one 
of the $\c$ field is exponentially decaying. Therefore, we can integrate out the $\c$ field, by proceeding as described
in detail in \cite[Section 3.1]{GGM}, and we get the analogue of \cite[Eq.(3.1)]{GGM}:
\be \Xi_{--}(\L_{\ell,L})=e^{L\ell E_0}\int P(d\psi)e^{\VV^{(0)}(\psi)} \label{3.fs}\ee
where $E_0$ is an analytic function of $\l$, vanishing at $\l=0$ and 
weakly depending on the volume; namely, 
if we denote by $E^\io_0$ its infinite volume limit, then $|E_0-E_0^\io|\le C|\l| e^{-c\ell}$ for two suitable constants 
$C,c>0$; see below for a more detailed discussion of the finite volume corrections to the thermodynamic
quantities of the theory. 
\vskip.2truecm
{\it Multiscale analysis.}
We now need to integrate out the $\psi$ field. However, since the $\psi$ field is massless, we cannot do it 
trivially in one step. A convenient way to proceed is by an iterative procedure, described in detail in 
\cite[Section 3.2]{GGM}. 
We define a sequence of geometrically decreasing momentum scales
$2^h$, with $h=1,0,\ldots$ Correspondingly we define a sequence of cutoff functions $f_h(\kk)$ 
in the following way. Let $\c(t)$ be a smooth compact
support function that is $1$ for $t\le 1$ and $0$ for $t\ge 2$.
We choose $f_0(\kk)=1-\c(|\kk|)$ and $f_h(\kk')=\c(2^{-h}|\kk|)
-\c(2^{-h+1}|\kk|)$
$\forall h<0$, so that $f_h$ for $h<0$ is non zero only if $2^{h-1}\le |\kk|\le 2^{h+1}$, and 
\be 1=\sum_{h\le 0} f_h(\kk)\;.\label{3.7}\ee
The resolution of the identity Eq.(\ref{3.7}) induces a rewriting
of the propagator of $\psi=: \psi^{(\le 0)}$ as a sum of
propagators concentrated on smaller and smaller momentum scales
and an iterative procedure to compute $Z$. At each step we
decompose the propagator into a sum of two propagators, the first
approximately supported on momenta $\sim 2^{h}$ (i.e.\ with a
Fourier transform proportional to $f_h(\kk)$), $h\le 0$, the
second approximately supported on momenta smaller than $2^h$.
Correspondingly we rewrite the Grassmann field as a sum
of two independent fields: $\psi^{(\le h)}=\psi^{(h)}+ \psi^{(\le
h-1)}$ and we integrate out the field $\psi^{(h)}$ in the same way
as we did for $\c$. The result is that, for any $h\le 0$, we can
rewrite
\be \Xi_{--}(\L_{\ell,L})=e^{L\ell E_h}\int P(d\psi^{(\le h)})e^{\VV^{(h)}(\psi^{(\le h)})} \label{h20}
\ee
where $E_h, \VV^{(h)}$ are defined recursively, and $P(d\psi^{(\le h)})$
is the gaussian integration with propagator given in momentum space by 
$\sum_{h'\le h}\hat g^{(h')}(\kk)$, with 
\be \hat g^{(h)}(\kk)=\frac{2\p}{Z}f_h(\kk)\big[C_\psi(\kk) \big]^{-1}\;.\label{3.gh}\ee
Note that the direct space counterpart of $g^{(h)}$ decays to zero faster than any power:
\be |g^{(h)}(\xx)|\le \frac{C_p}{(1+2^h|\boldsymbol{\d}(\xx)|)^p}\;,\quad \forall p\ge 0\;.\label{3.gbound}\ee
where $$\boldsymbol{\d}(\xx)=\Big(\frac\ell\p\sin\big(\frac{\p x_1}\ell\big),\frac{L}\p\sin\big(\frac{\p x_2}{L}\big)\Big)\;.$$
The outcome of the iterative construction is that, in particular, the 
 {\it effective potential} $\VV^{(h)}$ has the following structure: 
\be \VV^{(h)}(\psi)=\sum_{n\ge 1}\sum_{{\underline\o},{\underline \xx}}\,  W^{(h)}_{2n;{\underline\o}}
({\underline\xx})\Big[\prod_{i=1}^{2n}\psi_{\xx_i,\o_i}\Big]\;,
\label{3.5q}\ee
where $\underline\xx=(\xx_1,\ldots,\xx_{2n})$, $\underline\o=(\o_1,\ldots,\o_{2n})$ and
\be  \frac1{L\ell}\sum_{\xx_1,\ldots,\xx_{2n}}
|W^{(h)}_{2n;{\underline\o}}(\xx_1,\ldots,\xx_{2n})|\le C^{n}2^{h (2-n)} |\l|^{\max\{1,c\,n\}}\label{3.5}\ee
for suitable $C,c>0$. For future reference, let us also rewrite Eq.(\ref{3.5q}) in momentum space:
\be \VV^{(h)}(\psi)=\sum_{n\ge 1}\frac1{(L\ell)^{2n-1}}\!\!\sum_{\substack{\kk_1,\ldots,\kk_{2n}\\
{\underline\o}}}\!\!\hat W^{(h)}_{2n;{\underline\o}}
(\kk_1,\ldots,\kk_{2n-1})\Big[\prod_{i=1}^{2n}\hat \psi_{\kk_i,\o_i}\Big]\d(\kk_1+ \cdots +\kk_{2n})
\label{3.5qmom}\ee
where $\d(\kk)$ is a Kronecker's delta, periodic over the torus $\mathbb{R}^2/2\p\mathbb{Z}^2$.

The iteration continues until the scale $h^*:=\lfloor \log_2 (\p/\ell)\rfloor$ is reached. At that point,
the left-over propagator, $g^{(\le h^*)}$ is massive on the ``right scale" (i.e. on the very same scale
$2^{h^*}$), so that the associated degrees of freedom can be integrated in one step. The result is the 
desired partition function. 
\vskip.2truecm
{\it Localization and renormalization.}
In order to inductively prove Eq.(\ref{h20}) we write
\be {\cal V}^{(h)}(\psi) =\LL{\cal V}^{(h)}(\psi)+\RR{\cal
V}^{(h)}(\psi)\;,\label{3.loc} \ee
where $\LL{\cal V}^{(h)}(\psi)$ and $\RR{\cal V}^{(h)}(\psi)$ are the so-called {\it local} and {\it irrelevant}
part of the effective potential, defined in the next few formulas.
We use a definition of localization operator on the lattice at finite
volume, analogous to the one used in \cite{BM01} where the finite volume effects of the renormalization procedure 
are discussed in great detail. To be fair, the definitions below are a bit more complicated than those in 
\cite{BM01} (see the slightly cumbersome definitions (\ref{3.37})-(\ref{3.38})), the reason being that we want to make sure that the relative finite size errors induced by the 
localization procedure are of the order $O(\ell^{-4})$ rather than $O(\ell^{-2})$, which would not be enough to our purposes; see Eqs.(\ref{3.40})--(\ref{3.42})
and (\ref{3.err}) below. Let us now come back to the definition of $\LL{\cal V}^{(h)}(\psi)$ and $\RR{\cal V}^{(h)}(\psi)$. If we think of the kernel $\hat W^{(h)}_{2,0;(\o_1,\o_2)}$ as a $2\times2$ matrix with 
matrix indices $\o_1,\o_2$, we let 
\bea 
&&\LL{\cal V}^{(h)}(\psi)=\frac1{L\ell}\sum_{\kk}\hat \psi_{\kk}^T
\big[\LL \hat W_{2}^{(h)}(\kk)\big]
\hat\psi_{-\kk}
\label{2.localize}\\ &&\quad  +\frac1{(L\ell)^3}\sum_{\kk_1,\kk_2,\kk_3}\big[\LL\hat W^{(h)}_{4;\underline{\o}}(\kk_1,\kk_2,\kk_3)\big]
\hat \psi^{(\le h)}_{\kk_1,\o_1}\hat \psi^{(\le h)}_{\kk_2,\o_2}\hat \psi^{(\le h)}_{\kk_3,\o_3}
\hat \psi^{(\le h)}_{-\kk_1-\kk_2-\kk_3,\o_4}
\;,\nonumber\eea
where, setting $\bar\kk_{\h\h'}=(\h\frac{\p}{\ell},\h'\frac{\p}{L})$,
\bea &&\LL \hat W_{2}^{(h)}(\kk)=\frac14\sum_{\h,\h'=\pm}\Biggl[\frac98
\hat W_{2}^{(h)}(\bar\kk_{\h\h'})
\Big(1+\h\frac{\sin k_1}{\sin (\p/\ell)}+\h'\frac{\sin k_2}{\sin(\p/L)}\Big)\nonumber\\
&&\quad \quad-\frac18
\hat W_{2}^{(h)}(3\bar\kk_{\h\h'})
\Big(1+\h\frac{\sin k_1}{\sin (3\p/\ell)}+\h'\frac{\sin k_2}{\sin(3\p/L)}\Big)\Biggr]\;,\label{3.37}\eea
and
\bea &&\LL\hat W^{(h)}_{4;\underline{\o}}(\kk_1,\kk_2,\kk_3)=
\frac1{64}\sum_{\substack{\h_1,\h_2,\h_3,\\ \h'_1,\h'_2,\h'_3}}
\Biggl[\frac98 \hat W^{(h)}_{4;\underline{\o}}(\bar\kk_{\h_1\h'_1},
\bar\kk_{\h_2\h'_2},\bar\kk_{\h_3\h'_3})\nonumber\\
&&\quad -\frac18
\hat W^{(h)}_{4;\underline{\o}}(3\bar\kk_{\h_1\h'_1},
3\bar\kk_{\h_2\h'_2},3\bar\kk_{\h_3\h'_3})\Biggr]\;.\label{3.38}\eea
Note that in the limit $L,\ell\to\infty$, the action of the localization operator reduces to:
$$\LL \hat W_{2}^{(h)}(\kk)=\hat W_{2}^{(h)}(\V0)
+(\sin k_1\dpr_{k_1}+\sin k_2\dpr_{k_2})\hat W_{2}^{(h)}(\V0)$$ and, similarly, 
$$\LL\hat W^{(h)}_{4;\underline{\o}}(\kk_1,\kk_2,\kk_3)=\hat W^{(h)}_{4;\underline{\o}}(\V0,\V0,\V0)\;.$$ 
In other words, at finite $L$ and $\ell$, $\LL \hat W_{2}^{(h)}(\kk)$ and $\LL\hat W^{(h)}_{4;\underline{\o}}(\kk_1,\kk_2,\kk_3)$ 
have to be understood as finite volume, lattice versions of the Taylor expansion up to order $0$ or $1$, respectively. 
The coefficients $9/8$ and $-1/8$ have been fixed in such a way that the difference 
between the finite and infinite volume localization operators goes to zero as $[\max\{L,\ell\}]^{-4}$;
if needed, we could further modify the definition of localization in such a way that the operator remains the 
same in the infinite volume limit, but the difference with its finite volume counterpart goes to zero faster than any power 
of $\max\{L,\ell\}$. Note also that $\LL$ has the nice feature of being a projection operator: $\LL^2=\LL$. 

The direct-space counterparts of Eqs.(\ref{3.37})-(\ref{3.38}) read as follows:
\bea &&\LL\sum_{\xx,\yy}\psi_\xx^{(\le h)} W_{2}^{(h)}(\xx,\yy)\psi_\yy^{(\le h)}=\label{3.37sp}\\
&&=\sum_{\xx,\yy}\psi_\xx^{(\le h)}W_{2}^{(h)}(\xx,\yy)\Big[G_{\yy,\xx}\psi_\xx^{(\le h)}+\sum_{i=1}^2
d_i(\yy,\xx)\bar\dpr_i\psi^{(\le h)}_\xx\Big]
\;,\nonumber\eea
and
\bea&& \LL\sum_{\substack{\xx_1,\xx_2,\xx_3,\xx_4\\ \o_1,\o_2,\o_3,\o_4}}
 W^{(h)}_{4;\underline{\o}}(\xx_1,\xx_2,\xx_3,\xx_4)\prod_{i=1}^4\psi^{(\le h)}_{\xx_i,\o_i}=\label{3.38sp}\\
&&= \sum_{\substack{\xx_1,\xx_2,\xx_3,\xx_4\\ \o_1,\o_2,\o_3,\o_4}}
W^{(h)}_{4;\underline{\o}}(\xx_1,\xx_2,\xx_3,\xx_4)\prod_{i=1}^4\Big[G_{\xx_i,\xx_4}\psi^{(\le h)}_{\xx_4,\o_i}\Big]
\;,\nonumber\eea
where $G_{\yy,\xx}$ and $d_i(\yy,\xx)$ are translation invariant and  
\bea && G_{\xx,\V0}=\frac98\cos\big(\frac{\p x_1}{\ell}\big)\cos\big(\frac{\p x_2}{L}\big)
-\frac18\cos\big(\frac{3\p x_1}{\ell}\big)\cos\big(\frac{3\p x_2}{L}\big)\;,\label{3.40}\\
&& d_1(\xx,\V0)=\frac98\frac{\sin\big(\frac{\p x_1}{\ell}\big)}{\sin\big(\frac\p\ell\big)}
\cos\big(\frac{\p x_2}{L}\big)
-\frac18\frac{\sin\big(\frac{3\p x_1}{\ell}\big)}{\sin\big(\frac{3\p}\ell\big)}
\cos\big(\frac{3\p x_2}{L}\big)\;,\label{3.41}\\
&& d_2(\xx,\V0)=\frac98\cos\big(\frac{\p x_1}{\ell}\big)
\frac{\sin\big(\frac{\p x_2}{L}\big)}{\sin\big(\frac\p{L}\big)}
-\frac18\cos\big(\frac{3\p x_1}{\ell}\big)
\frac{\sin\big(\frac{3\p x_2}{L}\big)}{\sin\big(\frac{3\p}L\big)}\label{3.42}
\;,\eea
and $\bar\dpr_i$ is the symmetric discrete derivative w.r.t. $x_i$, i.e., $\bar\dpr_1 f(x_1,x_2)=
\frac12\big[f(x_1+1,x_2)-f(x_1-1,x_2)\big]$, and similarly for $\bar\dpr_2$. A few comments are now in order.\\
1)
The action of $\LL$ in direct space can be seen as an action on the fields, as indicated by 
Eqs.(\ref{3.37sp})-(\ref{3.38sp}). The action of $\RR=1-\LL$ can be interpreted in the same way, too: it takes the form 
\bea && \hskip-.5truecm\RR\sum_{\xx,\yy}\psi_\xx^{(\le h)} W_{2}^{(h)}(\xx,\yy)\psi_\yy^{(\le h)}=\sum_{\xx,\yy}\psi_\xx^{(\le h)} W_{2}^{(h)}(\xx,\yy)T_{\yy,\xx}^{(\le h)}\label{3.43}\\
&& \hskip-.5truecm\RR\sum_{\underline{\xx},\underline{\o}}
 W^{(h)}_{4;\underline{\o}}(\underline{\xx})
\prod_{i=1}^4\psi^{(\le h)}_{\xx_i,\o_i}
 =\sum_{\underline{\xx},\underline{\o}}
 W^{(h)}_{4;\underline{\o}}(\underline{\xx})\Big[D^{(\le h)}_{\xx_1,\xx_4;\o_1}\psi^{(\le h)}_{\xx_2,\o_2}
 \psi^{(\le h)}_{\xx_3,\o_3}\psi^{(\le h)}_{\xx_4,\o_4}+\nonumber\\
 &&\hskip-.5truecm+G_{\xx_1,\xx_4}\psi^{(\le h)}_{\xx_4,\o_1}D^{(\le h)}_{\xx_2,\xx_4;\o_2} \psi^{(\le h)}_{\xx_3,\o_3}\psi^{(\le h)}_{\xx_4,\o_4}+
G_{\xx_1,\xx_4}\psi^{(\le h)}_{\xx_4,\o_1}G_{\xx_2,\xx_4}\psi^{(\le h)}_{\xx_4,\o_2}D^{(\le h)}_{\xx_3,\xx_4;\o_3} 
\psi^{(\le h)}_{\xx_4,\o_4}  \Big]\nonumber
 \eea
where 
\bea && T_{\yy,\xx;\o}^{(\le h)}=\psi^{(\le h)}_{\yy,\o}-
G_{\yy,\xx}\psi_\xx^{(\le h)}-\sum_{i=1}^2
d_i(\yy,\xx)\bar\dpr_i\psi^{(\le h)}_\xx\;,\\
&& D_{\yy,\xx;\o}^{(\le h)}=\psi^{(\le h)}_{\yy,\o}-
G_{\yy,\xx}\psi_\xx^{(\le h)}\eea
\noindent
2) The functions  $G_{\yy,\xx}$ and $d_i(\yy,\xx)$ are antiperiodic over $\L_{\ell,L}$ in both their arguments. 
This implies that the fields $T_{\yy,\xx;\o}^{(\le h)}$ and $D_{\yy,\xx;\o}^{(\le h)}$ are antiperiodic in $\yy$ and 
periodic in $\xx$. Therefore, the summands in Eqs.(\ref{3.37sp})-(\ref{3.38sp}) are periodic both in $\yy$ and $\xx$.
\\
3) For fixed $\xx$, asymptotically as $L\gg\ell\gg|\xx|$, we can rewrite
\be
G_{\xx,\V0}=1+O\big((|\xx|/\ell)^4\big)\;,\qquad d_i(\xx,\V0)=x_i\big[1+O\big((|\xx|/\ell)^4\big)\big]\;.\label{3.err}\ee
which is useful for bounding dimensionally the effect of the finite volume on the localization procedure. \\
4) The action of $\RR$ produces a dimensional gain on the Grassmann monomials which it acts on non-trivially. 
This can be seen as follows. 
As discussed in item (1), the action of $\RR$ can be thought of as a replacement of a field 
$\psi^{(\le )}_\yy$ either by $T^{(\le h)}_{\yy,\xx}$ or by $D^{(\le h)}_{\yy,\xx}$, with 
$\xx,\yy\in\L_{\ell,L}$ (recall that the box is topologically a torus); since both $\xx$ and $\yy$ 
are represented 
by infinitely many equivalent images on $\mathbb{Z}^2$, it is always possible to pick two such images,
to be called $\bar\xx,\bar\yy$, so that their euclidean distance on $\mathbb{Z}^2$ is the same as the
distance between $\xx$ and $\yy$ on the torus $\L_{\ell,L}$. We have 
$D_{\yy,\xx}^{(\le h)}=(-1)^{(y_1-\bar y_1)/\ell+(y_2-\bar y_2)/L}D_{\bar\yy,\bar\xx}^{(\le h)}$,
where 
$D_{\bar\yy,\bar\xx}^{(\le h)}=
\psi^{(\le h)}_{\bar\yy}-
G_{\bar\yy,\bar\xx}\psi_{\bar \xx}^{(\le h)}$ can be conveniently written as:
\be D_{\bar\yy,\bar\xx;\o}^{(\le h)}=
(\bar\yy-\bar\xx)\cdot \int_0^1 ds \dpr_{\xx'}\psi^{(\le h)}_{\xx'}\big|_{
\xx'=\xx+s(\yy-\xx)}+(1-G_{\bar\yy,\bar\xx})\psi^{(\le h)}_{\bar\xx}\label{3.blah}\ee
where $(1-G_{\bar\yy,\bar\xx})$ is of the order $|\bar\yy-\bar\xx|^4\ell^{-4}$. 
Note that when we plug this formula in the last two lines of Eq.(\ref{3.43}), 
the factors $(\bar\yy-\bar\xx)$ or $(1-G_{\bar\yy,\bar\xx})$ multiply the kernel $W_4^{(h)}$, whose value is obtained 
by integrating the degrees of freedom on scales strictly larger than $h$, i.e., it is a sum 
over Feynman diagrams with propagators $g^{(h_i)}(\xx_i-\yy_i)$ all of scale $h_i>h$. Consider the contribution 
from the interpolated term first. When we decompose
the factor $(\bar\yy-\bar\xx)$ along the  diagrams contributing to $W^{(h)}_4$, some of the propagators $g^{(h_i)}(\xx_i-\yy_i)$ are multiplied
by $(\xx_i-\yy_i)$, which is dimensionally equivalent to a factor $2^{-h_i}$ (see Eq.(\ref{3.gbound})).
On the other hand, the derivative 
$\dpr_{\xx'}$ inside the interpolation integral, when acting on $\psi^{(\le h)}_{\xx'}$, 
is dimensionally equivalent to a factor $2^h$, simply because $\psi^{(\le h)}_{\xx'}$
is a field that is smooth on scale $2^{-h}$. Similarly, the contribution from the second term in Eq.(\ref{3.blah})
can be bounded by first replacing $(1-G_{\bar\yy,\bar\xx})$ by $({\rm const.})|\bar\yy-\bar\xx|^4\ell^{-4}$
and then by decomposing the factor $|\bar\yy-\bar\xx|^4$ along the diagrams contributing to $W^{(h)}_4$,
so obtaining $2^{-4h_i}\ell^{-4}=
2^{4(h^*-h_i)}$ for some $h_i>h$. In conclusion, the action of $\RR$ on the four-legged kernels 
is dimensionally equivalent to a dimensional gain
$2^{h-h_i}$, with $h_i>h$, if we pick the first term in Eq.(\ref{3.blah}), or to $2^{4(h^*-h_i)}$ if we pick the second term, which 
is due to the finite volume corrections to the localization procedure.
A similar discussion applies to the action of $\RR$ on the two-legged kernels, in which case $\RR$ 
is equivalent to either $2^{2(h-h_i)}$, with $h_i>h$, if we pick the analogue of the first term 
in Eq.(\ref{3.blah}), or to $2^{4(h^*-h_i)}$, if we pick the analogue of the second term.\\
5) A key fact which makes the theory at hand treatable (and asymptotically free)
is that the action of $\LL$ on the quartic kernels is zero ``by the Pauli principle", i.e., simply by the Grassmann 
rule $\psi_{\xx,\o}^2=0$. In fact, note that in the second line of Eq.(\ref{3.38sp})
at least two of the four $\o$ indices must be equal among each other. 
Therefore the expression in the second line of Eq.(\ref{2.localize}) is identically zero. This property can be diagramatically interpreted by 
saying that the fermionic nature of the theory automatically renormalizes the 
four-field interaction, which is dimensionally marginal (see below) 
but effectively irrelevant thanks to the cancellation that we just mentioned.\\
6) By the previous comment, the only non vanishing contribution to the local part of the effective action is the one
in the first line of Eq.(\ref{2.localize}). By its very definition, it is apparent that $\LL\hat W^{(h)}_2$ is invariant under the discrete symmetries of the theory (reflections, discrete rotations, etc, see \cite[Section II.D]{GGM} for
a thorough discussion of this point). Therefore, its most general form is (see \cite[(2.68)]{GGM})
\be \LL \hat W_2^{(h)}(\kk)=\begin{pmatrix}z_h( i\sin k_1-\sin k_2 )& i2^h\n_h \\ -i2^h\n_h & z_h( i\sin k_1+\sin k_2)\end{pmatrix}
\ee
for two {\it real} constants $z_h,\n_h$, which will be called the {\it running coupling constants} of our theory. 
The initial values of these constants $z_0,\n_0$, are induced by the choices of $t_c,Z$ in Eq.(\ref{3.18}), as well as
by the effects of the change of variables described after Eq.(\ref{3.21}) and of the integration of the massive fields in 
Eq.(\ref{3.fs}). It is straightforward (if lengthy) to keep track of this series of transformations and to check that  
$z_0,\n_0$ are analytically invertible functions of $t_c, Z$ in a neighborood of $t_c=t_c^0, Z=1,\l=0$:
\be (z_0,\n_0)=\big(F_0(\l,t_c,Z),N_0(\l,t_c,Z)\big)\quad \Leftrightarrow\quad \begin{cases}
Z=1+\z(\l,z_0,\n_0)\\ t_c=t_c^0+\t(\l,z_0,\n_0)\end{cases}\label{3.50}\ee
where $F_0,N_0,\z,\t$ are analytic functions of their arguments. It is straightforward (if lengthy) to check that,
if $|\l|+|\n_0|+|z_0|\le \e_0$,  
\bea && \frac{\dpr \z(\l,z_0,\n_0)}{\dpr z_0}=a+O(\e_0)\;,\qquad \frac{\dpr \z(\l,z_0,\n_0)}{\dpr \n_0}=O(\e_0)\;,
\label{3.51}\\
&& \frac{\dpr \t(\l,z_0,\n_0)}{\dpr z_0}=O(\e_0)\;,\qquad \frac{\dpr \t(\l,z_0,\n_0)}{\dpr \n_0}=b+O(\e_0)\;,\eea
for two {\it non vanishing} constants $a$ and $b$. This property 
will play an important role in the following, in the choice of the parameter $Z$.  
\vskip.2truecm
{\it Tree expansion.} Going back to the inductive proof of Eq.(\ref{h20}), we note that representation is valid at the 
first step, see Eq.(\ref{3.fs}). Assuming the representation to be valid at scale $h$, let us show that the same
structure is preserved at the following step. By using the rewriting Eq.(\ref{3.loc})
and the {\it addition principle} (see e.g. \cite[Section 4]{GeM}),
we rewrite Eq.(\ref{h20}) as
\be e^{L\ell E_h}\int P(d\psi^{(\le h-1)})\int P(d\psi^{(h)})
e^{\LL\VV^{(h)}(\psi^{(\le h-1)}+\psi^{(h)})+\RR\VV^{(h)}(\psi^{(\le h-1)}+\psi^{(h)})}\ee
where $P(d\psi^{(h)})$
is the gaussian integration with propagator $\hat g^{(h)}(\kk)$, see Eq.(\ref{3.gh}). By integrating out 
the degrees of freedom on scale $h$, which are massive with mass of the order of $2^h$, we get:
\bea && e^{L\ell (E_h+e_h)}\int P(d\psi^{(\le h-1)})
e^{\VV^{(h-1)}(\psi^{(\le h-1)})}\;,\label{wd}\\
&&L\ell e_h+\VV^{(h-1)}(\psi)=\log \int P(d\psi^{(h)})
e^{\LL\VV^{(h)}(\psi+\psi^{(h)})+\RR\VV^{(h)}(\psi+\psi^{(h)})}\;,\nonumber
\eea
which proves the inductive hypothesis Eq.(\ref{h20}), provided that $E_{h-1}$ is fixed as
\be E_{h-1}=E_h+e_h=E_0+\sum_{h\le j\le 0}e_j\;.\ee
Note that the above procedure allows us to write the
running coupling constants $z_h,\n_h$ with $h\le 0$,
in terms of $z_k,\n_k$ with $h<k\le 0$:
\be z_{h-1}=z_h+\b^{z}_h\;,\qquad \n_{h-1}=2\n_h+\b^\n_h\;,
\label{pp} \ee
where $\b_h^\#=\b_h^\#\big((z_h,\n_h),\ldots,
(z_0,\n_0)\big)$ is the so--called {\it Beta function}.

By the very definition of {\it truncated expectation} (see e.g. \cite[(2.57)]{GGM}), the second line of Eq.(\ref{wd}) can 
be rewritten as 
\bea&& L\ell e_h+\VV^{(h-1)}(\psi)=\\
&&=\sum_{n\ge 1}
\EE^T_h( \underbrace{\LL \VV^{(h)}(\psi+\cdot)\!+\!\RR \VV^{(h)}(\psi+\cdot);
\cdots;\LL \VV^{(h)}(\psi+\cdot)\!+\!\RR \VV^{(h)}(\psi+\cdot)}_{n\ {\rm times}})\nonumber\eea
where $\EE^T_h$ is the truncated expectation associated with the gaussian integration $P(d\psi^{(h)})$.
Iterating this relation, we are led to a tree expansion for the effective potential, as described in detail in e.g.
 \cite[Sections III.A- III.D]{GGM}. The resulting trees are defined in a way that is very similar to those in 
 \cite[Sections III.D]{GGM}, with the following minor differences (we refer for comparison to the 
 description of trees in items 1 to 5 in \cite[Section III.D]{GGM}, to be called GGM.1-GGM.5, 
 as well as to the discussion following that 
 item list):\\
1) The trees we consider have only normal endpoints: $m=0$ in the notation of item GGM.1.\\
2) The ultraviolet scale $N$ (in the notation of GGM.2) is replaced by $0$ and we denote by 
${\cal T}^{(h)}_{n}$ the set of labeled trees with root on scale $h$ and $n$ endpoints.\\
3) The first four lines of GGM.5 are replaced by ``With each normal endpoint $v$ on scale $2$ we associate a factor
$\lis\VV(\psi^{(\le 0)},\c)$; here $\c:=\psi^{(1)}$ should be thought of as
the field on scale $1$. With the endpoints on scale $h_v\le 1$ we associate either a factor
$2^h\n_h\frac{1}{L\ell}\sum_{\kk,\o}i\o\hat \psi_{\kk,\o}
\hat\psi_{-\kk,-\o}$ or $z_h\frac{1}{L\ell}\sum_{\kk,\o}\hat \psi_{\kk,\o}(i\sin k_1-\o
\sin k_2)
\hat\psi_{-\kk,\o}$, in which cases we shall refer to the endpoint as being of type $\n$ or $z$, respectively."
\vskip.2truecm
In terms of the definitions of trees, 
the effective potential 
${\cal V}^{(h)}$ can be written,  in analogy with \cite[(3.44)]{GGM} as
\be L\ell e_h+{\cal V}^{(h)}(\psi^{(\le h)})  =
\sum_{\substack{n\ge 1}}\sum_{\t\in\TT^{(h)}_{n}}
\VV^{(h)}(\t,\psi^{(\le h)})\;,\label{GGM3.17}\ee
where, if $v_0$ is the first vertex of $\t$, if $\t_1,\ldots,\t_s$ ($s=s_{v_0}$)
are the subtrees of $\t$ with root $v_0$, and if $\EE^T_{h+1}$ is the truncated expectation 
associated to the propagator $\hat g^{(h)}(\kk)$,
\be {\cal V}^{(h)}(\t,\psi^{(\le h)})=\frac{1}{s!} \EE^T_{h+1}
\big(\lis V^{(h+1)}(\t_1,\psi^{(\le h+1)});\ldots; \lis V^{(h+1)}
(\t_{s},\psi^{(\le h+1)})\big)\;,\label{GGM3.18}\ee
and $\lis V^{(h+1)}(\t_i,\psi^{(\le h+1)})$:
\begin{itemize}
\item is equal to $\RR\VV^{(h+1)}(\t_i,\psi^{(\le h+1)}))$ if $\t_i$ is non trivial;
\item is equal to $2^{h+1}\n_{h+1}\frac{1}{L\ell}\sum_{\kk,\o}i\o\hat \psi_{\kk,\o}
\hat\psi_{-\kk,-\o}$ if $\t_i$ is trivial, $h< N$ and the endpoint
of $\t_i$ is of type $\n$;
\item is equal to $z_{h+1}\frac{1}{L\ell}\sum_{\kk,\o}\hat \psi_{\kk,\o}(i\sin k_1-\o\sin k_2)
\hat\psi_{-\kk,\o}$ if $\t_i$ is trivial, $h< N$ and the endpoint
of $\t_i$ is of type $z$;
\item is equal to $\lis \VV(\psi^{(\le N+1)})$
if $\t_i$ is trivial and $h=0$.
\end{itemize}
The values of the trees can be estimated dimensionally as described in detail in 
\cite[Sections III.D.1-III.D.2]{GGM}. In a notation analogous to the one introduced in those sections, we get 
the analogue of \cite[(3.35)]{GGM}, namely, if $W_{\t,{\bf P},{\bf T},\b}$ is the kernel of the renormalized 
effective potential labelled by a tree $\t\in{\cal T}^{(h)}_n$, a set of field labels ${\bf P}$, a spanning tree $T$ 
and a set of interpolation parameters $\b$, 
\bea && \frac1{L\ell}\sum_{\b\in B_T}\sum_{\xx_{v_0}} |W_{\t,{\bf P},T,\b}(\xx_{v_0})|\le
C^{n}\Big[\prod_{v\ {\rm e.p.}}(C|\l|)^{\max\{1,c|P_{v}|\}}\Big]\cdot\nonumber\\
&&\cdot 2^{h(2-\frac12|P_{v_0}|)}\Big[\prod_{v\,{\rm not}\ {\rm e.p.}} 2^{(h_v-h_{v'})(2-\frac12|P_v|-z(P_v))}
\Big]\;,\label{5.51tv}\eea
with $z(P_v)=2\d_{|P_v|,2}+\d_{|P_v|,4}$. 
\vskip.2truecm
{\bf Remark.} The bound Eq.(\ref{5.51tv}) is essentially dimensional, i.e., it can be understood (modulo the 
combinatorics) by replacing all the propagators and the integrations involved in the definition 
of the kernel by their dimensional estimates, that is $|g^{(k)}(\xx)|\le ({\rm const.})2^{k}$ and 
$\int d\xx|g^{(k)}(\xx)|\le ({\rm const.})2^{-k}$, see Eq.(\ref{3.gbound}). To be fair, the dimensional estimate that we would
obtain by these simple replacements would be similar to Eq.(\ref{5.51tv}), but with $z(P_v)$ replaced by zero. 
In order to justify the presence of the dimensional factors $\prod_{v\,{\rm not}\ {\rm e.p.}} 2^{-(h_v-h_{v'})z(P_v)}$,
one needs to take into account the action of $\RR$, see the discussion in item (4) above in the lines preceding and following Eq.(\ref{3.blah}).
It is important to note that in order for the bound in Eq.(\ref{5.51tv}) to be summable over the scale labels, 
it is not really necessary to have $z(P_v)=2\d_{|P_v|,2}+\d_{|P_v|,4}$: a gaining factor 
$\tilde z(P_v)=(1+\e)\d_{|P_v|,2}+\e\d_{|P_v|,4}$ with $\e\in(0,1)$ would make the job, too. 
For this reason, in order to regularize the kernels it is actually enough to use a portion of the gaining factors described in item (4) above; if desired, we can keep some of them on a side. E.g., concerning 
the factors $2^{4(h^*-h_i)}$ coming from the finite volume
corrections to the definition of $\RR$, we can use a fraction $2^{(1+\e)(h^*-h_i)}$ with $\e\in(0,1)$ to regularize the kernels, and 
we can keep $2^{(3-\e)(h^*-h_i)}$ on a side.
\vskip.2truecm
The bound Eq.(\ref{5.51tv}) is valid 
{\it provided that the running coupling constants remain of order $\l$}, for all scales between $h$ and $0$.
Under this assumption, Eq.(\ref{5.51tv})  implies the analyticity of the 
kernels of $\VV^{(h)}$ and the decay bounds Eq.(\ref{3.5}).

An immediate corollary of the bound Eq.(\ref{5.51tv}) is
that contributions from trees $\t\in\TT_{n}^{(h)}$ with a vertex $v$ on
scale $h_v=k>h$ admit an improved bound with respect to
Eq.(\ref{3.5}), with an extra dimensional factor $2^{\th(h-k)}$, $0<\th<1$,
which can be thought of as a dimensional gain with
respect to the ``basic'' dimensional bound in Eq.(\ref{3.5}). This
improved bound is usually referred to as the {\it short memory}
property (i.e., long trees are exponentially suppressed); it is due 
to the fact that the renormalized scaling dimensions $d_v=2-\frac12|P_v|-z(P_v)$
in Eq.(\ref{5.51tv})  are all $\le -1$, and can be obtained by taking a fraction of 
the factors $2^{(h_v-h_{v'})d_v}$ associated to the branches of the tree
$\t$ on the path connecting the vertex on scale $k$ to the one on scale $h$.

Under the same assumptions, the beta function itself, $\b_h^\#$, is analytic and dimensionally bounded 
by a constant independent of $h$. Moreover, the contributions to it from trees that have at least one node 
on scale $k>h$ is dimensional bounded proportionally to $2^{\th(h-k)}$, with $0<\th<1$.
It is remarkable that thanks to these bounds, the dynamical system 
induced by the beta function can be fully studied and shown to lead to a bounded 
flow of the running coupling constants.
\vskip.4truecm
{\it The flow of the running coupling constants.}
As announced above, the flow equations for the running coupling constants are
\be z_{h-1}=z_h+\b^{z}_h\;,\qquad \n_{h-1}=2\n_h+\b^\n_h\;,\ee
where $\b_h^\#=\b_h^\#\big((z_h,\n_h),\ldots,
(z_0,\n_0)\big)$ is an analytic function of its argument, with an analyticity 
domain bounded by: $|z_k|+|\n_k|\le \e_0$, for all 
$h\le k\le 0$, where $\e_0$ is a suitable (small) positive constant. 

Note that both $\b_h^z$ and $\b_h^\n$ can be expressed as sums over trees with at least 
one endpoint on scale $2$, the reason being that the local part of the trees with only 
endpoints of scale $\le 1$ is zero by the support properties of the single-scale propagators
that enter the definition of $\b_h^\#$.
Therefore, by the short memory property, $|\b_h^z|,|\b_h^\n|\le C_\th|\l|2^{\th h}$, for $\th\in(0,1)$, uniformly 
in $L,\ell$. The idea is to first 
solve the flow equations in the $L,\ell\to\infty$ limit, by properly choosing the initial data in such a way that 
the sequence $\{(z_h,\n_h)\}_{h\le 0}$ remains bounded. Then we will use the same initial data as the $L,\ell=\io$ case in the finite volume equations and we will show that the resulting flow remains bounded and close to the
infinite volume one, with explicit bounds on the error terms. 

Let us then consider the case $L,\ell=\infty$ first. We denote by $\b_h^{\io,\#}$ the corresponding beta function. 
We define 
$\mathfrak{M}_{K,\th}$ to be the space of sequences
$\underline v=(\underline z, \underline \n)=\{(z_h,\n_h)\}_{h\le 0}$ such that $|z_h|+|\n_h|\le
K|\l|2^{\th h} $, $\forall h\le 0$; we shall think of
$\mathfrak{M}_{K,\th}$ as a Banach space with norm
$||\cdot||_\th$, where $||\underline v||_{\th}=\sup_{k\le
0}(|z_h|+|\n_h|)2^{-\th k}$. Note that every exponentially decaying solution to the 
beta function equations (if any) can be looked for as a fixed point of the map $\mathbf{T}:
\mathfrak{M}_{K,\th} \to \mathfrak{M}_{K,\th} $ defined by
\be (\mathbf{T}
\underline z)_h = -\sum_{ j\le h}
\b_{j}^{\io,z}(\underline v)\;,\qquad 
(\mathbf{T} \underline \n)_h = -\sum_{ j\le h} 2^{j-h-1}
\b_{j}^{\io,\n}(\underline v)\;.\label{3.map}\ee
The fact that, for $K$ sufficiently large, $\mathbf{T} $ is a map
from $\mathfrak{M}_{N;K,\th}$ to itself is a simple consequence of
the bound $|\b_h^{\io,\#}|\le C_\th|\l|2^{\th h}$. Moreover, if $\underline v ,\underline v'\in\mathfrak{M}_{K,\th}$, then
using the short memory property:
\be |\b_{j}^{\io,\#}(\underline v)-\b_j^{\io,\#}(\underline v')|\le
C_\th'|\l|2^{j\th}\sum_{k\ge j}|v_k-v_k'|\le C''_\th|\l|2^{j\th}
||\underline v-\underline v'||_\th\;,\label{3.58}\ee
which implies that
\be ||\mathbf{T} \underline v -  \mathbf{T} \underline v '
||_{\th}
 \le C'''_\th |\l| \| \underline v -  \underline v ' \|_{\th}\;,\ee
i.e.\ $\mathbf{T}$ is a contraction for $|\l|$ sufficiently
small.  Then the Banach fixed point theorem implies that $\mathbf{T}$ has a
unique fixed point 
$\underline{v}^*$ in  $\mathfrak{M}_{K,\th}$, which represents an exponentially decaying solution 
to the flow equations, with initial data of order $\l$ and given explicitly by the following expressions:
\be z_0^*=z_0^*(\l,Z):=-\sum_{ j\le 0}  \b_j^{\io,z}(\underline v^*)\;,\qquad 
\n_0^*=\n_0^*(\l,Z):=-\sum_{ j\le 0} 2^{j-1} \b_j^{\io,\n}(\underline v^*)\;.\label{3.58s}\ee
By the previous construction, 
the functions $z_0^*(\l,Z)$ and $\n_0^*(\l,Z)$ are analytic in their arguments in a neighborood
of $\l=0$, $Z=1$. 

We now consider the case of finite $L,\ell$ and we pick the same initial datum 
$(z_0,\n_0)$ for the flow equation as for the infinite volume case:
\be z_0=z_0^*(\l,Z)\;,\qquad \n_0=\n_0^*(\l,Z)\;.\label{3.63}\ee
Denoting by $\bar{\underline v}=\{(\bar z_h,\bar \n_h)\}_{h^*\le h\le 0}$ the sequence of running coupling constants generated at finite volume 
by this initial datum (using the notation above, the infinite volume counterpart of this sequence
is denoted by $\underline v^*=\{(z_h^*,\n_h^*)\}_{h\le 0}$), we get 
\be \begin{cases}
\bar z_h-z_h^*=\sum_{h< j\le 0}[\b_{j}^{z}(\bar{\underline v})-
\b_{j}^{\io,z}(\underline v^*)]\;,\\
\bar \n_h-\n_h^\io=\sum_{h< j\le 0}2^{j-h-1}[\b_{j}^{\n}(\bar{\underline v})-
\b_{j}^{\io,\n}(\underline v^*)]\;.\end{cases}\label{3.barz}\ee
Based on this equation, and thinking of $\bar{\underline v}$ as an infinite sequence (obtained e.g. by posing 
$\bar v_h=0$, $\forall h<h^*$), we can prove that, for any $\th<\e<1$,
\be
|\bar z_h-z_h^*|+|\bar\n_h-\n_h^*|\le C_{\th,\e} 2^{\th h}\Big(\frac{2^{-h}}{\ell}\Big)^{3-\e}
\label{3.wec}\;.\ee
The proof is by induction in $h$. If $h=0$ the claim is obviously true, simply because the l.h.s. is 
zero. For $h<0$, assuming the estimate to be valid for all the scales $h<k\le 0$, we use Eq.(\ref{3.barz}), 
by rewriting the expressions in square brackets as
\be
[\b_{j}^{\#}(\bar{\underline v})-
\b_{j}^{\io,\#}(\bar{\underline v})]+[\b_{j}^{\io,\#}(\bar{\underline v})-
\b_{j}^{\io,\#}(\underline v^*)]\;.\label{3.64}
\ee
Now, the second term can be bounded as in Eq.(\ref{3.58}), $|\b_{j}^{\io,\#}(\bar{\underline v})-
\b_{j}^{\io,\#}(\underline v^*)|\le C_\th'|\l|2^{j\th}\sum_{k\ge j}|\bar v_k-v_k^*|$, 
so that by using the inductive assumption we get that the corresponding contribution, $\sum_{h<j\le 0}
|\b_{j}^{\io,\#}(\bar{\underline v})-\b_{j}^{\io,\#}(\underline v^*)|$, is bounded from above 
by $C''_\th |\l|2^{2h\th}\big(2^{-h}\ell^{-1}\big)^{3-\e}$, as desired. 

The first term in Eq.(\ref{3.64}) is due to the finite volume corrections. Remember that, by using 
the tree construction explained above, $\b_{j}^{\#}(\bar{\underline v})$ can be written as a
sum of the form $\frac1{L\ell}\sum_{\t,\a}\sum_{\xx_1,\ldots,\xx_{n_\a}}\tilde W_j^{\#}(\xx_1,\ldots,
\xx_{n_\a})$, where $\a$ is a suitable multi-index (collecting the indices ${\bf P},T,\b$ indicated in Eq.(\ref{5.51tv}))
and $\tilde W$ is periodic over $\L_{\ell,L}$ in all its coordinates. By construction, $\tilde W_j^\#(\underline{\xx})$
is a combination of propagators on the scales indexed by the tree labels, as well as of the functions $G,d_i$ 
in Eqs.(\ref{3.40})--(\ref{3.42})
resulting from the action of $\RR$ on the nodes of the tree. Moreover
$\tilde W_j^\#(\xx_1,\ldots,\xx_{n_\a})$ is translation invariant, so that we can fix
one variable to $\V0$, for instance $\xx_1$, and write:
\bea && \b_{j}^{\#}(\bar{\underline v})-
\b_{j}^{\io,\#}(\bar{\underline v})=\label{3.fin}\\
&&=
\sum_{\t,\a}\Big[\sum_{\xx_2,\ldots,\xx_{n_\a}\in\L_{\ell,L}}\tilde W_j^{\#}(\V0,\xx_2\ldots,\xx_{n_\a})-\hskip-.6truecm
\sum_{\xx_2,\ldots,\xx_{n_\a}\in\mathbb Z^2}\tilde W_j^{\io,\#}(\V0,\xx_2\ldots,\xx_{n_\a})\Big]
\;,\nonumber\eea
where $\tilde W_j^{\io,\#}$ is the infinite 
volume limit of $\tilde W_j^{\#}$, which differs from the latter because of the replacement 
of  the factors $G$ and $d_i$ by $1$ and $x_i$, respectively, and for the replacement 
of the propagators $g^{(k)}$ by their infinite volume limit. 

Now, the two sums $\sum_{\xx_2,\ldots,\xx_{n_\a}}$ 
in Eq.(\ref{3.fin}) can be written as $\sum_{\xx_2,\ldots,\xx_{n_\a}}^*+$ $\sum_{\xx_2,\ldots,\xx_{n_\a}}^{**}$,
where $\sum_{\xx_2,\ldots,\xx_{n_\a}}^*$ is over the set $|\xx_i|\le \ell/4$, $\forall i=2,\ldots, n_\a$, while 
$\sum_{\xx_2,\ldots,\xx_{n_\a}}^{**}$ involves at least one coordinate outside the ball $B_{\ell/4}=\{\xx:|\xx|\le \ell/4\}$.
The easiest terms to bound are $$\sum_{\xx_2,\ldots,\xx_{n_\a}}^{**}\big[|\tilde W_j^{\#}(\V0,\xx_2\ldots,\xx_{n_\a})|+
|\tilde W_j^{\io,\#}(\V0,\xx_2\ldots,\xx_{n_\a})|\big]\;,$$ which are of the order $|\l|2^{j\th} (2^{-j}/\ell)^p$ for an arbitrary 
$p\ge 0$, simply because $\tilde W_j^{\#}$ contains a chain of propagators (each decaying 
faster than any power on a scale $h_i\ge j$, see Eq.(\ref{3.gbound})) connecting $\V0$ with a coordinate
$\bar\xx$ outside the ball $B_{\ell,4}$. We are left with bounding 
\be \sum_{\t,\a}\sum_{\xx_2,\ldots,\xx_{n_\a}\in B_{\ell/4}}
\big[\tilde W_j^{\#}(\V0,\xx_2\ldots,\xx_{n_\a})-\tilde W_j^{\io,\#}(\V0,\xx_2\ldots,\xx_{n_\a})\big]\;,\ee
where the differences in square brackets can be written as a sum of terms each of which involves either the 
difference between $G$ and $1$ (or, similarly, $d_i$ and $x_i$), or the difference between a propagator 
$g^{(k)}$ and its infinite volume limit. Regarding the first class of terms, remember that 
the relative difference between 
$G,d_i$ and their infinite volume limit is bounded dimensionally by  
$2^{4(h^*-h_i)}$, see the discussion in item (4) above, in particular after Eq.(\ref{3.blah}). 
Part of this factor (at least a portion $2^{(1+\e)(h^*-h_i)}$) is needed in order to renormalize the kernels of the effective 
potential, see Remark following Eq.(\ref{5.51tv}). What we are left with is exactly a factor $2^{(3-\e)(h^*-h_i)}$, as 
commented at the end of that Remark.
Summing these contributions over the scales $j> h$ gives the desired bound Eq.(\ref{3.wec}). 
Finally, regarding the terms involving a difference between the finite and infinite volume propagators, we 
observe that by the Poisson 
summation formula, the finite volume propagator $g^{(h)}(\xx)$, which is antiperiodic in its argument, can 
be written as a sum over images: 
\be g^{(h)}(x_1,x_2)=\sum_{{\bf n}\in \mathbb{Z}^2}(-1)^{n_1+n_2}
g^{(h)}_{\io}(x_1+n_1\ell,x_2+n_2 L)=:g^{(h)}_\io(\xx)+
\d g^{(h)}(\xx)\;,\ee
where $\d g^{(h)}(\xx)$ is smaller than any power of $(2^{-h}/\ell)$, namely, if $|\xx|\le \ell/2$,
$|\d g^{(h)}(\xx)|\le C_p 2^h (2^{-h}/\ell)^p$, $\forall p\ge 0$. Therefore, the terms 
involving a difference between a finite and an infinite volume propagator are smaller than any power in $(2^{-j}/\ell)$,
which is more than enough to the purpose of deriving the desired bound
Eq.(\ref{3.wec}). This concludes the proof of that bound. 
\vskip.3truecm
In order to complete the discussion related to the choice of the initial data $z_0,\n_0$, we are left with 
inverting the relation for $Z$, which is obtained by combining Eq.(\ref{3.63}) with the infinite volume limit of 
Eq.(\ref{3.50}), namely:
\be Z=1+\z\big(\l,z_0^*(\l,Z),\n_0^*(\l,Z)\big)\;.\label{3.70}\ee
The key ingredients to be used are the derivative estimates Eq.(\ref{3.51}) together 
with the observation that the derivatives with respect to $Z$ of the propagators obey to the same 
decay bound as the propagators themselves, so that 
\be
\frac{\dpr z_0^*}{\dpr Z}=-\sum_{j\le 0}  \frac{\dpr \b_j^{\io,z}(\underline v^*)}{\dpr Z}
 =O(\l)\;,
\ee
simply because $\partial_Z \b_j^{\io,z}(\underline v^*)=O(\l 2^{j \th'})$ for some $\th'<\th$. 
Therefore we can apply the implicit 
function theorem to invert Eq.(\ref{3.70}). This concludes the discussion about the choice of the 
initial data for the flow equation and, correspondingly, of the parameter $Z$ in Eq.(\ref{3.11}). 
Note that once $Z$ is fixed, the critical temperature $t_c$ is given by Eq.(\ref{3.50}):
\be t_c=t_c^0+\t\big(\l,z_0^*(\l,Z),\n_0^*(\l,Z)\big)\;.\label{3.70bis}\ee

\subsection{Explicit computation of the bulk and finite volume corrections
to the pressure}
\label{subsec3.4}

In this section we compute the bulk and correction terms from the free energy at the critical temperature, on the basis of 
Eqs.(\ref{fio})-(\ref{fiode}) and of the construction of $\mathcal Z_{--}(\L_{\ell,L})$ in the previous sections. 
Using Eq.(\ref{3.14}), we write
\be \log \mathcal Z_{--}(\L_{\ell,L})=\log C_{\ell,L}+\log \tilde {\mathcal Z}^0_{--}(\L_{\ell,L})+\log 
\int P(d\Phi)\, e^{\lis{\mathcal V}(\Phi)}\ee
and we compute separately the contributions from the three terms. 
\vskip.3truecm
{\it The term $\log C_{\ell,L}$.}
Using Eq.(\ref{2.52}) and the properties of the potential $v(\xx)$ spelled after Eq.(\ref{2.1}), 
as well as the definition of $V_{\ell,L}(\l)$ in Eq.(\ref{forV}), 
we can write
\bea  \frac1{\ell L}\log 
C_{\ell,L}&=&\log \big(2\cosh^2(\b_c J)\big)+\sum_{\xx\in\mathbb Z^2}\log\cosh\!\big(\frac{\b_c\l}2 v(\xx)\big)
+\nonumber\\
&+&2\sum_{\substack{\G\subseteq\L_{\ell,L}:\\
\supp\G\ni b_0}} \frac{\f^T(\G)}
{|\supp\G|} \prod_{\g\in\G} \z(\emptyset,\emptyset;\g)\;,\label{forVbis}\eea
where $\b_c$ is the interacting critical temperature (such that $t_c=\tanh(\b_c J)$, with $t_c$ fixed as in Eq.(\ref{3.70bis})) and $b_0$ can be chosen aribitrarily, e.g., it can be fixed to be the bond connecting $\V0$ with ${\bf e}_1$.
Note that the last term in the first line is independent of $\ell,L$, because $v(\xx)$ has finite range, while the 
term in the second line differs from its infinite volume limit by exponentially small terms, which 
correspond to the contributions from multipolygons $\G$ that either wind up over the torus $\L_{\ell,L}$ or 
touch the complement of $\L_{\ell,L}$ on $\mathbb Z^2$ (their exponential smallness follows from the decay bound 
Eq.(\ref{eq:zDef})). In conclusion, the term $\log C_{\ell,L}$ contributes to the bulk term of the free energy
Eq.(\ref{fio}),
\bea &&f_{\io;1}:=\lim_{\ell,L\to\infty}\frac1{\ell L}\log 
C_{\ell,L}=\log \big(2\cosh^2(\b_c J)\big)+\\
&&\qquad +\sum_{\xx\in\mathbb Z^2}\log\cosh\!\big(\frac{\b_c\l}2 v(\xx)\big)
+2\sum_{\substack{\G\subseteq\mathbb Z^2:\\
\supp\G\ni b_0}} \frac{\f^T(\G)}
{|\supp\G|} \prod_{\g\in\G} \z(\emptyset,\emptyset;\g)\;,\nonumber\eea
but not to the finite volume correction Eq.(\ref{fiode}).
\vskip.3truecm
{\it The term $\log \tilde {\mathcal Z}^0_{--}(\L_{\ell,L})$.}
By its very definition, see the lines preceding Eq.(\ref{3.14}), we can write
\be \frac1{\ell L}\log \tilde {\mathcal Z}^0_{--}(\L_{\ell,L})=\frac1{\ell L}\log \int \mathcal D\Phi e^{Z S_{t_c^0}(\Phi)}=
2\log Z+\frac1{\ell L}\log \int \mathcal D\Phi e^{S_{t_c^0}(\Phi)}\;,\ee
where in the last identity we performed the Grassmann change of variables $\Phi\to\sqrt{Z}\Phi$. 
The last term, $\frac1{\ell L}\log \int \mathcal D\Phi e^{S_{t_c^0}(\Phi)}$, is the non interacting pressure 
evaluated at the critical point (up to an additive constant $\frac1{\ell L}\log C_{\ell,L}\big|_{\l=0,\b=\b_c^0}=
\sqrt2+1$), whose bulk and finite volume corrections have been evaluated in great detail 
in Section \ref{sec22}. Putting things together we get:
\bea &&f_{\io;2}:= \lim_{\ell,L\to\infty}\frac1{\ell L}\log \tilde {\mathcal Z}^0_{--}(\L_{\ell,L})=\\
&&\qquad=\log \frac{\sqrt2 Z^2}{\sqrt2+1}
+\frac12\int\limits_{[-\p,\p]^2}\!\!\!\frac{d\kk}{(2\p)^2}\log(4-2\cos k_1-2\cos k_2)
\nonumber\eea
and 
\be \lim_{\ell\to\infty}\lim_{L\to\infty}\Big[\frac{\ell}{L}\log \tilde {\mathcal Z}^0_{--}(\L_{\ell,L})-\ell^2f_{\io;2}\Big]=\frac\p{12}\;.\ee
\vskip.3truecm
{\it The term $\log 
\int P(d\Phi)\, e^{\lis{\mathcal V}(\Phi)}$.} By the Renormalization Group analysis of $\int P(d\Phi)\, e^{\lis{\mathcal V}(\Phi)}$ described above, we can write:
\be \frac1{\ell L}\log 
\int P(d\Phi)\, e^{\lis{\mathcal V}(\Phi)}=E_{h^*}=E_0+\sum_{h^*\le h\le 0}e_h\ee
with $E_0$ and $e_h$ defined by Eqs.(\ref{3.fs}) and (\ref{wd}). Using the fact that
$|z_k|+|\n_k|\le C |\l| 2^{\th k}$ and the tree expansion explained above, 
we can bound
\be |e_h|\le C |\l| 2^{2h} 2^{\frac{\th}2 h}\;, \label{3.84}\ee
uniformly in $\ell,L$, for all $\th\in(0,1)$. Denoting by $E_0^\io$ and $e^\io_h$ the infinite volume limits of $E_0,e_h$, we can write the contribution to the bulk free energy under consideration as:
\be f_{\io;3}:=\lim_{\ell,L\to\infty}\log 
\int P(d\Phi)\, e^{\lis{\mathcal V}(\Phi)}=E_0^\io+\sum_{h\le 0}e_h^\io\;,\ee
which is an exponentially convergent series, whose sum is of order $\l$. The finite volume correction of interest
can then be written as
\bea && \lim_{\ell\to\infty}\lim_{L\to\io}\Big[\frac{\ell}{L}\log\int P(d\Phi)\, e^{\lis{\mathcal V}(\Phi)}
-\ell^2f_{\io;2}\Big]=\label{3.86}\\
&&\qquad = \lim_{\ell\to\infty}\ell^2\big[(\tilde E_0-E_0^\io)+\sum_{h^*\le h\le 0}(\tilde e_h-e_h^\io)-\sum_{h<h^*}
e_h^\io\big]\;,\nonumber\eea
where $\tilde E_0=\lim_{L\to\infty}E_0$ and $\tilde e_h=\lim_{L\to\infty}e_h$. Now, 
$(\tilde E_0-E_0^\io)$ is exponentially small in $\ell$, as it follows from the fact that it can be written as a sum 
of terms that involve at least one difference between the finite and infinite volume propagator of the $\c$ field,
which is exponentially small in $\ell$; therefore $\lim_{\ell\to\infty}\ell^2(\tilde E_0-E_0^\io)=0$. 
The last contribution to the r.h.s. of Eq.(\ref{3.86}) is also easy to estimate; in fact, using Eq.(\ref{3.84})
and the definition of $h^*=\lfloor\log_2 (\p/\ell)\rfloor$, we get:
\be \sum_{h<h^*}
|e_h^\io|\le ({\rm const.})|\l|2^{2h^*} 2^{\frac{\th}2 h^*}\le ({\rm const.})|\l|\ell^{-2-\frac{\th}{2}}\;,
\ee
which implies that $ \lim_{\ell\to\infty}\ell^2\sum_{h<h^*}
e_h^\io=0$. We are left with
\be  \lim_{\ell\to\infty}\ell^2\sum_{h^*\le h\le 0}(\tilde e_h-e_h^\io)\;.\label{3.88}\ee
Remember that $e_h$ (as well as $\tilde e_h$ and $e_h^\io$)
is a function of the whole sequence of coupling constants $\bar {\underline{v}}$ 
(see the lines following Eq.(\ref{3.63}) for a definition of $\bar{ \underline{v}}$ and of its infinite volume counterpart
$\underline{v}^*$);
we shall also indicate by $\tilde{\underline{v}}$ the $L\to\infty$ limit of $\bar{ \underline {v}}$. We can then write
Eq.(\ref{3.88}) as
\be  \lim_{\ell\to\infty}\ell^2\sum_{h^*\le h\le 0}\big[(\tilde e_h(\underline{\tilde v})-e_h^\io(\underline{\tilde v}))
+(e_h^\io(\underline{\tilde v})-e_h^\io(\underline{v}^*))
\big]\;.\label{3.89}\ee
Thanks to the bound Eq.(\ref{3.wec}) on the difference between the finite and infinite volume running coupling constants, we immediately see that the second term in square brackets can be bounded as:
\be |e_h^\io(\underline{\tilde v})-e_h^\io(\underline{v}^*)|\le ({\rm const.})|\l|2^{(2+\frac{\th}{2})h}\Big(\frac{2^{-h}}{\ell}\Big)^{3-\e}\;.\label{3.90}\ee
Picking $\th,\e>0$ small enough, we get %
\be \sum_{h^*\le h\le 0}|e_h^\io(\underline{\tilde v})-e_h^\io(\underline{v}^*)|
\le ({\rm const.})|\l|2^{(2+\frac{\th}{2})h^*}\Big(\frac{2^{-h^*}}{\ell}\Big)^{3-\e}
\le ({\rm const.})|\l|\ell^{-2-\frac{\th}{2}}\;,\ee
which implies $\lim_{\ell\to\io}\ell^2\big(e_h^\io(\underline{\tilde v})-e_h^\io(\underline{v}^*)\big)=0$. We are left 
with the first term in square brackets in Eq.(\ref{3.89}), which can be studied in a way similar to Eq.(\ref{3.fin}). 
By repeating a discussion completely analogous to the one following Eq.(\ref{3.fin}), we find that 
$|\tilde e_h(\underline{\tilde v})-e_h^\io(\underline{\tilde v})|$ admits the same bound 
Eq.(\ref{3.90}) as $ |e_h^\io(\underline{\tilde v})-e_h^\io(\underline{v}^*)|$, so that also
$\lim_{\ell\to\io}\ell^2\big(\tilde e_h(\underline{\tilde v})-e_h^\io(\underline{\tilde v})\big)=0$.
In conclusion, the finite volume corrections Eq.(\ref{3.86}) are exactly zero. Putting all the contributions together, 
we find that the central charge $c$ defined by Eq.(\ref{fiode}) is independent of $\l$ and equal to $1/2$, as desired. 

\section{Proof of the partition function inequality.}\label{sec4}

In this section we prove Lemma \ref{lemma1}. We start by proving it in the case $\l=0$ with generic couplings $\{J_b\}$.
In this case, using Eqs.(\ref{4.1k})--(\ref{4.4k}), we immediately find 
\be \mathcal{Z}^0_{+-}(\{J_b\};\L_{\ell,L})+\mathcal{Z}^0_{-+}(\{J_b\};\L_{\ell,L})=2
\mathcal{Z}^0_{e-e}(\{J_b\};\L_{\ell,L})+2\mathcal{Z}^0_{o-o}(\{J_b\};\L_{\ell,L})\label{3.kl}\ee
and
%
%
\bea &&\hskip-.8truecm \mathcal{Z}^0_{+-}(\{J_b\};\L_{\ell,L})+\mathcal{Z}^0_{-+}(\{J_b\};\L_{\ell,L})+\mathcal{Z}^0_{--}(\{J_b\};\L_{\ell,L})=3\mathcal{Z}^0_{e-e}(\{J_b\};\L_{\ell,L})+\nonumber\\
&&+
\mathcal{Z}^0_{e-o}(\{J_b\};\L_{\ell,L})+\mathcal{Z}^0_{o-e}(\{J_b\};\L_{\ell,L})+\mathcal{Z}^0_{o-o}(\{J_b\};\L_{\ell,L})
\;.\label{4.91}\eea
Remember that the four partition functions $\mathcal{Z}^0_{e-e}(\{J_b\};\L_{\ell,L})$, 
$\mathcal{Z}^0_{e-o}(\{J_b\};\L_{\ell,L})$,  $\mathcal{Z}^0_{o-e}(\{J_b\};\L_{\ell,L})$ and $\mathcal{Z}^0_{o-o}(\{J_b\};\L_{\ell,L})$ are all positive, see the definitions Eq. (\ref{h13z}) and following lines. Therefore, the right hand side of Eq.(\ref{3.kl}) is $\ge 0$, which proves  Eq.(\ref{3.12a}) for $\l=0$ and bond-dependent couplings. 
Moreover, the r.h.s. of Eq.(\ref{4.91}) is bounded from above by 
\bea && \hskip-.7truecm 3\mathcal{Z}^0_{e-e}(\{J_b\};\L_{\ell,L})+
3\mathcal{Z}^0_{e-o}(\{J_b\};\L_{\ell,L})+3\mathcal{Z}^0_{o-e}(\{J_b\};\L_{\ell,L})+3\mathcal{Z}^0_{o-o}(\{J_b\};\L_{\ell,L})\nonumber\\
&&\equiv3\mathcal{Z}^0(\{J_b\};\L_{\ell,L})\;,\eea
where we used Eq.(\ref{4.compk}). This proves the  lower bound in Eq.(\ref{3.11a})
for $\l=0$ and bond-dependent couplings. The upper bound in Eq.(\ref{3.11a}) for this case is proved analogously:
it is enough to observe that the r.h.s. of Eq.(\ref{4.91}) can be bounded from below by 
\bea && \mathcal{Z}^0_{e-e}(\{J_b\};\L_{\ell,L})+
\mathcal{Z}^0_{e-o}(\{J_b\};\L_{\ell,L})+\mathcal{Z}^0_{o-e}(\{J_b\};\L_{\ell,L})+\mathcal{Z}^0_{o-o}(\{J_b\};\L_{\ell,L})
\nonumber\\ &&\equiv\mathcal{Z}^0(\{J_b\};\L_{\ell,L})\;,\eea
which leads to the desired bound. 

\begin{figure}[h]
\centering
\includegraphics[width=0.8\textwidth]{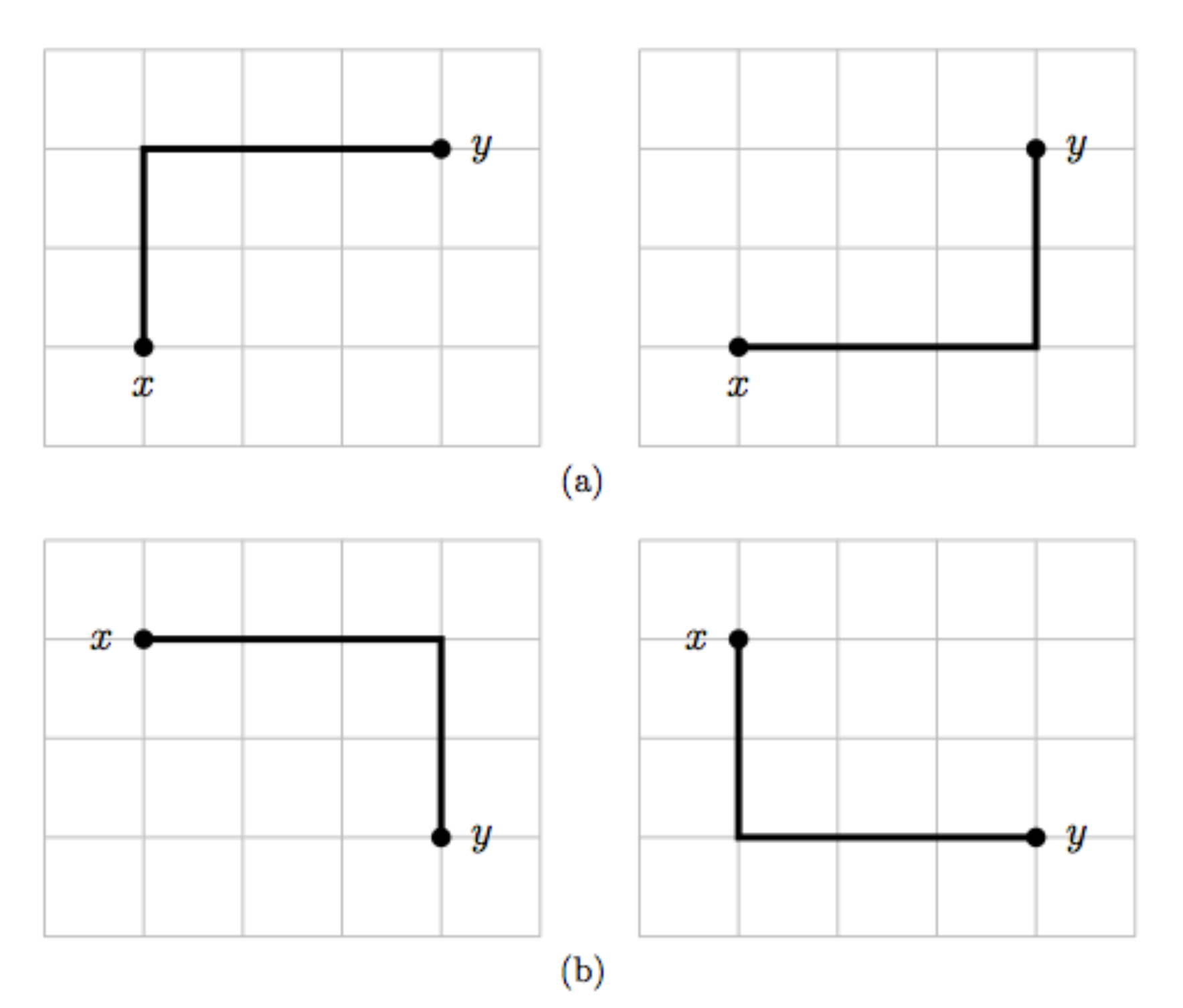}
\caption{The four possible types of ``corner like" strings}
\label{fig.corner}
\end{figure}

Let us now turn to the interacting case. The key issue is to obtain a representation of the four 
Grassmann partition functions $\mathcal{Z}_{++}(\L_{\ell,L})$, $\mathcal{Z}_{+-}(\L_{\ell,L})$, 
$\mathcal{Z}_{-+}(\L_{\ell,L})$, $\mathcal{Z}_{--}(\L_{\ell,L})$ in terms of (positive!) multipolygons partition sums. 
Such a representation is implicitly derived in \cite[Section II.A]{GGM}. Let us make it explicit here. 
The starting point is the
representation for the partition function in terms of disconnected
polymers (see \cite[Eqs.(2.8)-(2.9)-(2.10)]{GGM}):
\bea \mathcal{Z}(\L_{\ell,L})&=&\Big[\prod_{\{\xx,\yy\}} \cosh^2 \!\big(\tfrac12 \b\l
v(\xx-\yy)\big)\Big] \sum_{\underline{\s}}e^{\b
J\sum_{b}\tilde\s_b}\cdot\nonumber\\
&&\cdot\sum_{n\ge 0}\sum_{\{\tilde\g_1,\ldots,\tilde\g_n\}\subseteq\L_{\ell,L}}\f(\{\tilde\g_1,\ldots,\tilde\g_n\})
\prod_{\tilde\g\in\tilde\G}
z(\tilde\g)\;, \label{d.1} \eea
where
\be z(\tilde\g)=\sum_{\substack{\SS \text{ connected}: \\
\tilde\g(\SS)=\tilde\g}}\Big[\prod_{S\in\SS}\tanh(\tfrac12 \b\l
v_S)\Big]\Big[\prod_{b \in {\rm bl}(\SS)}
\tilde\s_b\Big]\label{d.2}\ee
and:\begin{itemize} \item $b$ indicates a nearest neighbor bond
and $\tilde\s_b$ is the bond spin, i.e. the product of the two
spins at the vertices of $b$; \item $\SS=\{S_1,\ldots,S_m\}$ is a
set of strings, where each string $S_i$ is the union of the bonds
in a finite lattice path that can be either horizontal, or
vertical, or ``corner-like", as in Figure \ref{fig.corner}; moreover, $v_S:=v(\xx-\yy)$, where $\xx$ and $\yy$ are the first and last points connected by the path $S$ on $\L_{\ell,L}$. We say
that a set of strings $\SS=\{S_1,\ldots,S_m\}$ is connected if,
given $1\le i_0<j_0\le m$, we can find a sequence
$(S_{i_0},S_{i_1},\ldots,S_{i_p}\=S_{j_0})$ such that $S_{i_l}\cap
S_{i_{l+1}}\neq \emptyset$. From a graphical point of view, every
connected component $\SS$ corresponds in a non-unique way to a
{\it polymer} $\tilde \g(\SS)$, i.e., a connected set of bonds.
 It is helpful to color the bonds in $\tilde \g(\SS)$ black or gray, depending
on whether the given bond belongs to an odd or even number of
strings in $\SS$, and denote the set of bonds thus colored black
by ${\rm bl}(\SS)$. See \cite[Fig.4]{GGM}.
 \item the sum $\sum_{\{\tilde\g_1,\ldots,\tilde\g_n\}}$
in the
r.h.s.\ runs over sets of polymers $\{\tilde\g_1,\ldots,\tilde\g_n\}$, 
such that each polymer is
contained in $\L_{\ell,L}$. Moreover, the function
$\f(\{\tilde\g_1,\ldots,\tilde\g_n\})$ implements the hard core condition,
that is $\f$ is equal to 1 if none of the polymers overlap, and 0
otherwise (here two polymers overlap if and only if they have at
least one bond in common); the term with $n=0$ should be
interpreted as 1.
\end{itemize}

Now, in order 
to obtain the Grassmann representation Eq.(\ref{aaaa11}) one can proceed as explained in 
\cite[Section II.A]{GGM}: the idea is simply to rewrite Eq.(\ref{d.1}) as
\bea && \mathcal{Z}(\L_{\ell,L})=\Big[\prod_{\{\xx,\yy\}} \cosh^2 \!\big(\tfrac12 \b\l
v(\xx-\yy)\big)\Big]\cdot\sum_{n\ge 0}\sum_{\{\tilde\g_1,\ldots,\tilde\g_n\}\subseteq\L_{\ell,L}}\f(\{\tilde\g_1,\ldots,\tilde\g_n\})\cdot\nonumber\\
&&\cdot\sum_{\substack{\SS_1,\ldots,\SS_n
\text{ connected}: \\ \tilde\g(\SS_i)=\tilde\g_i}}
\Big[\prod_{i=1}^n\prod_{S\in\mathcal{S}_i}\tanh(\tfrac12 \b\l v_S)\Big]\sum_{\underline{\s}}\prod_{b \in {\rm bl}(\underline\SS)}
\tilde\s_b\, e^{\b J\sum_b\tilde\s_b}\;,\eea
where $\underline{\SS}:=(\SS_1,\ldots,\SS_n)$, bl$(\underline{\SS})=\cup_{i=1}^n{\rm bl}(\SS_i)$ 
and  
\be \sum_{\underline{\s}}\prod_{b \in {\rm bl}(\underline{\SS})}
\tilde\s_b\, e^{\b J\sum_b\tilde\s_b}=\Big[\prod_{b \in {\rm bl}(\underline{\SS})}\frac1\b\frac{\dpr}{\dpr J_b}\Big]\mathcal{Z}^0(\{J_b\};\L_{\ell,L})\big|_{J_b\equiv J}\ee
and then to re-express $\mathcal{Z}^0(\{J_b\};\L_{\ell,L})$ as a sum of Grassmann integrals, via Eqs.(\ref{aaaa}) and (\ref{2.5tt}).
This leads to Eq.(\ref{aaaa11}), with 
\bea &&\hskip-.5truecm \mathcal{Z}_{\boldsymbol{\alpha}}(\L_{\ell,L})=\Big[\prod_{\{\xx,\yy\}} \cosh^2 \!\big(\tfrac12 \b\l
v(\xx-\yy)\big)\Big]\cdot\sum_{n\ge 0}\sum_{\{\tilde\g_1,\ldots,\tilde\g_n\}\subseteq\L_{\ell,L}}\f(\{\tilde\g_1,\ldots,\tilde\g_n\})\cdot\nonumber\\
&&\hskip-.6truecm\cdot\hskip-.8truecm\sum_{\substack{\SS_1,\ldots,\SS_n
\text{ connected}: \\ \tilde\g(\SS_i)=\tilde\g_i}}
\Big[\prod_{i=1}^n\prod_{S\in\mathcal{S}_i}\tanh(\tfrac12 \b\l v_S)\Big]
\Big[\prod_{b \in {\rm bl}(\underline{\SS})}\frac1\b\frac{\dpr}{\dpr J_b}\Big]\mathcal{Z}^0_{\boldsymbol{\alpha}}(\{J_b\};\L_{\ell,L})\big|_{J_b\equiv J}\;,\label{44.99}\eea
which is equivalent to 
\cite[Eq.(2.11)]{GGM} that, if further manipulated, implies the representation Eq.(\ref{prop1}).
On the other hand, if we plug Eqs.(\ref{4.1k})--(\ref{4.4k}) into Eq.(\ref{44.99}), we immediately get 
\bea && \mathcal{Z}_{++}(\L_{\ell,L})=\mathcal{Z}_{e-e}(\L_{\ell,L})-\mathcal{Z}_{e-o}(\L_{\ell,L})-\mathcal{Z}_{o-e}(\L_{\ell,L})-\mathcal{Z}_{o-o}(\L_{\ell,L})\;,\nonumber
\label{4.1bis}\\
&& \mathcal{Z}_{+-}(\L_{\ell,L})=
\mathcal{Z}_{e-e}(\L_{\ell,L})+\mathcal{Z}_{e-o}(\L_{\ell,L})-\mathcal{Z}_{o-e}(\L_{\ell,L})+\mathcal{Z}_{o-o}(\L_{\ell,L})\;,
\nonumber\\
 && \mathcal{Z}_{-+}(\L_{\ell,L})=
\mathcal{Z}_{e-e}(\L_{\ell,L})-\mathcal{Z}_{e-o}(\L_{\ell,L})+\mathcal{Z}_{o-e}(\L_{\ell,L})+\mathcal{Z}_{o-o}(\L_{\ell,L})\;,
\nonumber\\
&& \mathcal{Z}_{--}(\L_{\ell,L})=
\mathcal{Z}_{e-e}(\L_{\ell,L})+
\mathcal{Z}_{e-o}(\L_{\ell,L})+\mathcal{Z}_{o-e}(\L_{\ell,L})-\mathcal{Z}_{o-o}(\L_{\ell,L})\;,\nonumber\\
&&\label{ciao}\eea
with 
\bea &&\hskip-.5truecm \mathcal{Z}_{e-e}(\L_{\ell,L})=\Big[\prod_{\{\xx,\yy\}} \cosh^2 \!\big(\tfrac12 \b\l
v(\xx-\yy)\big)\Big]\cdot\sum_{n\ge 0}\sum_{\{\tilde\g_1,\ldots,\tilde\g_n\}\subseteq\L_{\ell,L}}\f(\{\tilde\g_1,\ldots,\tilde\g_n\})\cdot\nonumber\\
&&\hskip-.7truecm\cdot\hskip-.8truecm\sum_{\substack{\SS_1,\ldots,\SS_n
\text{ connected}: \\ \tilde\g(\SS_i)=\tilde\g_i}}
\Big[\prod_{i=1}^n\prod_{S\in\mathcal{S}_i}\tanh(\tfrac12 \b\l v_S)\Big]
\Big[\prod_{b \in {\rm bl}(\underline{\SS})}\frac1\b\frac{\dpr}{\dpr J_b}\Big]\mathcal{Z}^0_{e-e}(\{J_b\};\L_{\ell,L})\big|_{J_b\equiv J}\;,\label{44.990}\eea
and analogously for the three other partition functions, with the label $e-e$ replaced at both sides by $e-o$, $o-e$, 
$o-o$,
respectively. If we use the definition Eq.(\ref{h13z}), we can rewrite the last factor in Eq.(\ref{44.990}) as:
\bea && \Big[\prod_{b \in {\rm bl}(\underline{\SS})}\frac1\b\frac{\dpr}{\dpr J_b}\Big]\mathcal Z^0_{e-e}(\{J_b\};\L_{\ell,L})\big|_{J_b\equiv J}=\nonumber\\
&&=2^{\ell L}t^{| {\rm bl}(\underline{\SS})|}\big[\prod_{b\in \mathcal B_{\ell,L}}\cosh(\b J_b)\big] \sum_{\G\subseteq \L_{\ell,L}}\nolimits^{(e-e)}\prod_{\g\in\G} \prod_{b\in \g}t_b(\underline{\SS})\;,\label{h13zz}
\eea
where
$t_{b}(\underline{\SS})$ is equal
either to $t$, if $b\not\in {\rm Bl}(\underline{\SS})=\cup_{\SS_i\in \underline{\SS}}{\rm
bl}(\SS_i)$, or to
$1/t$, if $b\in {\rm Bl}(\underline{\SS})$. An equivalent way of rewriting Eq.(\ref{h13zz}) is 
\be  \Big[\prod_{b \in {\rm bl}(\underline{\SS})}\frac1\b\frac{\dpr}{\dpr J_b}\Big]\mathcal Z^0_{e-e}(\{J_b\};\L_{\ell,L})\big|_{J_b\equiv J}
=t^{| {\rm bl}(\underline{\SS})|}\mathcal Z^0_{e-e}(\{\bar J_b(\underline{\SS})\};\L_{\ell,L})\;,\ee
where $\bar J_{b}(\underline{\SS})$ is equal
either to $J$, if $b\not\in {\rm Bl}(\underline{\SS})=\cup_{\SS_i\in \underline{\SS}}{\rm
bl}(\SS_i)$, or to 
$\b^{-1}{\rm arctanh}(1/t)$, if $b\in {\rm Bl}(\underline{\SS})$. Plugging this back into Eq.(\ref{44.990}) gives
\bea &&\hskip-.5truecm \mathcal{Z}_{e-e}(\L_{\ell,L})=\Big[\prod_{\{\xx,\yy\}} \cosh^2 \!\big(\tfrac12 \b\l
v(\xx-\yy)\big)\Big]\cdot\sum_{n\ge 0}\sum_{\{\tilde\g_1,\ldots,\tilde\g_n\}\subseteq\L_{\ell,L}}\f(\{\tilde\g_1,\ldots,\tilde\g_n\})\cdot\nonumber\\
&&\hskip-.0truecm\cdot\hskip-.8truecm\sum_{\substack{\SS_1,\ldots,\SS_n
\text{ connected}: \\ \tilde\g(\SS_i)=\tilde\g_i}}
\Big[\prod_{i=1}^n\prod_{S\in\mathcal{S}_i}\tanh(\tfrac12 \b\l v_S)\Big]
t^{| {\rm bl}(\underline{\SS})|}\mathcal Z^0_{e-e}(\{\bar J_b(\underline{\SS})\};\L_{\ell,L})\;,\label{4.13w}\eea
and analogous formulas are valid for the three other 
partition functions, with the label $e-e$ replaced at both sides by $e-o$, $o-e$, 
$o-o$,
respectively.

Now, the key observation is that if $\l v_S\ge 0$ for all $S$, then all the factors appearing in 
Eq.(\ref{4.13w}) (as well as in its analogues with $e-e$ replaced by $e-o$, $o-e$, 
$o-o$) are non-negative. Therefore, if we insert 
Eq.(\ref{4.13w}) and its analogues with
the label $e-e$ replaced by $e-o$, $o-e$, 
$o-o$, into Eq.(\ref{ciao}), and if we use the known bounds on the partition functions at $\l=0$ with 
bond-dependent couplings, proved at the beginning of this section, namely 
$$\hskip-.5truecm\mathcal Z^0(\{\bar J_b(\underline{\SS})\};\L_{\ell,L})\le \mathcal Z^0_{--}(\{\bar J_b(\underline{\SS})\};\L_{\ell,L})+\mathcal Z^0_{-+}(\{\bar J_b(\underline{\SS})\};\L_{\ell,L})+\mathcal Z^0_{+-}(\{\bar J_b(\underline{\SS})\};\L_{\ell,L})$$
$$\hskip-.5truecm \mathcal Z^0_{--}(\{\bar J_b(\underline{\SS})\};\L_{\ell,L})+\mathcal Z^0_{-+}(\{\bar J_b(\underline{\SS})\};\L_{\ell,L})+\mathcal Z^0_{+-}(\{\bar J_b(\underline{\SS})\};\L_{\ell,L})
\le 3 \mathcal Z^0(\{\bar J_b(\underline{\SS})\};\L_{\ell,L})$$
$$
\mathcal Z^0_{-+}(\{\bar J_b(\underline{\SS})\};\L_{\ell,L})+\mathcal Z^0_{+-}(\{\bar J_b(\underline{\SS})\};\L_{\ell,L})\ge  0$$
then we finally obtain the desired estimates Eqs.(\ref{3.11a})-(\ref{3.12a}). This concludes the proof of Lemma \ref{lemma1} and, therefore, of Theorem \ref{thm1}.

\appendix
\section{The contribution to the pressure from the ratio of the Grassmann partition functions}
\label{appZ}

\subsection{The non-interacting case}\label{seca.1}

In this section we prove the vanishing of the limit in the second line of Eq.(\ref{sl}), 
\be \lim_{\ell\to\infty}\lim_{L\to\io}\frac{\ell}{L}\log \Big[\frac12
\big(1+\frac{\mathcal Z^0_{-+}(\L_{\ell,L})}{\mathcal Z^0_{--}(\L_{\ell,L})}+\frac{\mathcal Z^0_{+-}(\L_{\ell,L})}{\mathcal Z^0_{--}(\L_{\ell,L})}\big)\Big|_{\b=\b_c}
\Big]=0\;.\label{Asl}\ee
To this purpose, we rewrite the partition functions in Eq.(\ref{pc}) as 
\bea && 
\mathcal Z^0_{--}(\L_{\ell,L})\Big|_{\b=\b_c}=(\sqrt2)^{\ell L}
\prod_{r=0}^{\ell-1}\Big[\prod_{n=0}^{L-1}(2a_{2r+1}-z_{2n+1}-z_{2n+1}^{-1})\Big]^{1/2}\;,
\nonumber\\
&& \mathcal Z^0_{-+}(\L_{\ell,L})\Big|_{\b=\b_c}=(\sqrt2)^{\ell L}\prod_{r=0}^{\ell-1}\Big[\prod_{n=0}^{L-1}(2a_{2r+1}-z_{2n}-z_{2n}^{-1})\Big]^{1/2}\;,
\label{pc2}\\
&&\mathcal Z^0_{+-}(\L_{\ell,L})\Big|_{\b=\b_c}=(\sqrt2)^{\ell L}
\prod_{r=0}^{\ell-1}\Big[\prod_{n=0}^{L-1}(2a_{2r}-z_{2n+1}-z_{2n+1}^{-1})\Big]^{1/2}\;,
\nonumber\eea
where $a_p:=2-\cos(\frac{\p}{\ell}p)$ and $z_p:=e^{i\frac{\p}{L}p}$. The expressions in square brackets in the r.h.s of these equations can be further rewritten and put in the form used by 
\cite{FF}. Consider first the expression in square brackets appearing in the definition 
of $\mathcal Z^0_{--}(\L_{\ell,L})\Big|_{\b=\b_c}$. Note that $z_{2n+1}$ are the $L$ roots of $-1$ and, therefore, $\prod_{n=0}^{L-1}(z-z_{2n+1})=z^L+1$, $\forall z\in\mathbb{C}$. In particular, $\prod_{n=0}^{L-1}z_{2n+1}=(-1)^L$, so that 
\bea &&\prod_{n=0}^{L-1}(2a_{2r+1}-z_{2n+1}-z_{2n+1}^{-1})=\prod_{n=0}^{L-1}
(z_{2n+1}^2-2a_{2r+1}z_{2n+1}+1)=\nonumber\\
&&=\prod_{n=0}^{L-1}
(z_{2n+1}-a_{2r+1}^+)(z_{2n+1}-a_{2r+1}^-)=\big[(a_{2r+1}^+)^L+1\big]
\cdot\big[(a_{2r+1}^+)^L+1\big]
\;,\nonumber\eea
where $a^\pm_p=a_p\pm\sqrt{a_p^2-1}$. Choosing $\g_p\ge 0$ in such a way that 
$\cosh\g_p=a_p$, we can further simplify this into 
\be \prod_{n=0}^{L-1}(2a_{2r+1}-z_{2n+1}-z_{2n+1}^{-1})=(e^{L\g_{2r+1}}+1)(e^{-L\g_{2r+1}}+1)
=4\cosh^2\frac{L\g_{2r+1}}2\;.\ee
Plugging this back into the definition of $\mathcal Z^0_{--}(\L_{\ell,L})\Big|_{\b=\b_c}$ gives $$\mathcal Z^0_{--}(\L_{\ell,L})\Big|_{\b=\b_c}=(\sqrt2)^{\ell L}\prod_{r=0}^{\ell-1}2\cosh\frac{L\g_{2r+1}}2\;,$$
which is the same as \cite[Eq.(2.3)]{FF}. Actually, by exchanging the roles of $\ell$ and $L$ we can rewrite 
$\mathcal Z^0_{--}(\L_{\ell,L})\Big|_{\b=\b_c}$ in two equivalent ways:
\be \mathcal Z^0_{--}(\L_{\ell,L})\Big|_{\b=\b_c}=(\sqrt2)^{\ell L}\prod_{r=0}^{\ell-1}2\cosh\frac{L\g_{2r+1}}2=
(\sqrt2)^{\ell L}\prod_{n=0}^{L-1}2\cosh\frac{\ell\tilde\g_{2n+1}}2
\;,\ee
where $\tilde \g_p\ge 0$ is defined by the condition that $\cosh\tilde\g_p=2-\cos(\frac{\p}{L}p)$.
Proceeding exactly in the same way for $\mathcal Z^0_{-+}(\L_{\ell,L})\Big|_{\b=\b_c}, 
\mathcal Z^0_{+-}(\L_{\ell,L})\Big|_{\b=\b_c}$ gives:
\bea && \mathcal Z^0_{-+}(\L_{\ell,L})\Big|_{\b=\b_c}=(\sqrt2)^{\ell L}
\prod_{r=0}^{\ell-1}2\sinh\frac{L\g_{2r+1}}2=(\sqrt2)^{\ell L}\prod_{n=0}^{L-1}2\cosh\frac{\ell\tilde\g_{2n}}2\;,\nonumber\\
&& \mathcal Z^0_{+-}(\L_{\ell,L})\Big|_{\b=\b_c}=(\sqrt2)^{\ell L}
\prod_{r=0}^{\ell-1}2\cosh\frac{L\g_{2r}}2=(\sqrt2)^{\ell L}\prod_{n=0}^{L-1}2\sinh\frac{\ell\tilde\g_{2n+1}}2\;.
\nonumber\eea
By using these formulas we rewrite the l.h.s. of Eq.(\ref{Asl}) as
\be \lim_{\ell\to\infty}\lim_{L\to\io}\frac{\ell}{L}\log \Big[\frac12\big(1+\prod_{r=0}^{\ell-1}
\tanh\frac{L\g_{2r+1}}2+\prod_{n=0}^{L-1}
\tanh\frac{\ell\tilde\g_{2n+1}}2\big)\Big]\;.\label{Asl.1}\ee
Since $0\le \tanh x\le 1$ for $x\ge 0$, it is apparent that the argument of the logarithm 
in this equation is positive and smaller than $3/2$, which implies that the limit in Eq.(\ref{Asl.1})
is zero.

\subsection{The interacting case}\label{subsec3.5}
In order to bound the ratios $\frac{|\mathcal Z_{-+}(\L_{\ell,L})|}{\mathcal Z_{--}(\L_{\ell,L})}$ and 
$\frac{|\mathcal Z_{+-}(\L_{\ell,L})|}{\mathcal Z_{--}(\L_{\ell,L})}$ appearing in Eq.(\ref{3.10b}),
we compute $\mathcal Z_{-+}(\L_{\ell,L})$ and $\mathcal Z_{+-}(\L_{\ell,L})$ by a renormalization group 
construction analogous to the one used to analyze $\mathcal Z_{--}(\L_{\ell,L})$. Everything is the same,
with a few obvious changes induced by the different boundary conditions. The important fact that makes the 
renormalization group construction of $\mathcal Z_{-+}(\L_{\ell,L})$ and $\mathcal Z_{+-}(\L_{\ell,L})$
possible, is that $h^*$ is finite for both: in the first case $h^*=\lfloor\log_2(\p/\ell)\rfloor$, as for 
$\mathcal Z_{--}(\L_{\ell,L})$, while in the second
$h^*=\lfloor\log_2(\p/L)\rfloor$. Of course, the bulk contributions to the free energy are the same for all these 
partition functions, so that
\bea && \frac{|\mathcal Z_{-+}(\L_{\ell,L})|}{\mathcal Z_{--}(\L_{\ell,L})}\le \Big[
\frac{\mathcal Z^0_{-+}(\L_{\ell,L})}{\mathcal Z^0_{--}(\L_{\ell,L})}\Big]\Big|_{\b=\b_c}\hskip-.3truecm\cdot\ 
e^{\ell L R_1(\l)}\;,\\
&&  \frac{|\mathcal Z_{+-}(\L_{\ell,L})|}{\mathcal Z_{--}(\L_{\ell,L})}\le 
\Big[\frac{\mathcal Z^0_{+-}(\L_{\ell,L})}{\mathcal Z^0_{--}(\L_{\ell,L})}\Big]\Big|_{\b=\b_c}\hskip-.3truecm\cdot\ 
e^{\ell L R_2(\l)}\;,\eea
where $\mathcal Z^0_{--}(\L_{\ell,L}),\big|_{\b=\b_c},
\mathcal Z^0_{-+}(\L_{\ell,L})\big|_{\b=\b_c},\mathcal Z^0_{+-}(\L_{\ell,L})\big|_{\b=\b_c}$  are the non interacting partition functions at criticality defined in Eqs.(\ref{zzaa})-(\ref{pc}),
and $R_1,R_2$, according to be renormalization group analysis, are bounded by
$c|\l|\ell^{-2-\th'}$ for some $c>0$ and $0<\th'<1$. As proved in Appendix \ref{appZ}, the ratios 
$\Big[
\frac{\mathcal Z^0_{-+}(\L_{\ell,L})}{\mathcal Z^0_{--}(\L_{\ell,L})}\Big]\Big|_{\b=\b_c}$ and $\Big[
\frac{\mathcal Z^0_{+-}(\L_{\ell,L})}{\mathcal Z^0_{--}(\L_{\ell,L})}\Big]\Big|_{\b=\b_c}$ are positive and smaller than 1 and, therefore, 
\be \frac{|\mathcal Z_{-+}(\L_{\ell,L})|}{\mathcal Z_{--}(\L_{\ell,L})}
+\frac{|\mathcal Z_{+-}(\L_{\ell,L})|}{\mathcal Z_{--}(\L_{\ell,L})}\le 
e^{C|\l|\ell^{-1-\th'} L }\;,\ee
for a suitable constant $C>0$. Plugging this back into the l.h.s. of Eq.(\ref{3.10b}) gives
\be \lim_{\ell\to\infty}\lim_{L\to\io}\frac{\ell}{L}\log \Big(1+\frac{|\mathcal Z_{-+}(\L_{\ell,L})|}{\mathcal Z_{--}(\L_{\ell,L})}+\frac{|\mathcal Z_{+-}(\L_{\ell,L})|}{\mathcal Z_{--}(\L_{\ell,L})}\Big)\le 
\lim_{\ell\to\infty}\lim_{L\to\io}\frac{\ell}{L}C|\l|\frac{L}{\ell^{1+\th'}}\;,\label{3.10bwrt}\ee
which proves Eq.(\ref{3.10b}).\\

{\bf Acknowledgements.} The research leading to these results has received
funding from the European Research Council under the European Union's Seventh
Framework Programme ERC Starting Grant CoMBoS (grant agreement n$^o$ 239694). We would like to thank 
Rafael Greenblatt for some very inspiring discussions about the possible validity of a partition function inequality
similar to the one proved above. 


\end{document}